\documentclass[useAMS,usenatbib]{mn2e}
\usepackage{graphicx}

\newcommand{\ltsima} {$\; \buildrel < \over \sim \;$}
\newcommand{\gtsima} {$\; \buildrel > \over \sim \;$}
\newcommand{\lta} {\lower.5ex\hbox{\ltsima}}
\newcommand{\gta} {\lower.5ex\hbox{\gtsima}}

\newcommand{\beq}{\begin{equation}}
\newcommand{\eeq}{\end{equation}}
\newcommand{\ba}{\begin{array}}
\newcommand{\ea}{\end{array}}

\title[Observational evidence of dissipative photospheres in GRBs.
]{Observational evidence of dissipative photospheres in gamma-ray bursts}

\author[F. Ryde et al.]{Felix~Ryde$^{1,2}$\thanks{email: fryde@kth.se}, 
Asaf~Pe'er$^{3}$,
Tanja~Nymark$^{1,2}$,
Magnus~Axelsson$^{1,2}$, \newauthor 
Elena~Moretti$^{1,2}$,
Christoffer~Lundman$^{1,2}$, 
Milan~Battelino$^{1,2}$,  
\newauthor Elisabeth~Bissaldi$^{5}$,  
James~Chiang$^{6}$,  
Miranda~S.~Jackson$^{1,2}$,
Stefan~Larsson$^{4}$, \newauthor 
Francesco~Longo$^{7,8}$,   
Sinead~McGlynn$^{1,9}$,  
Nicola~Omodei$^{6}$, \\ 
$^{1}$Department of Physics, Royal Institute of Technology (KTH), AlbaNova, SE-106 91 Stockholm, Sweden\\ 
$^{2}$The Oskar Klein Centre for Cosmoparticle Physics, AlbaNova, SE-106 91 Stockholm, Sweden\\ 
$^{3}$Harvard-Smithsonian Center for Astrophysics, Cambridge, MA 02138, USA\\   
$^{4}$Department of Astronomy, Stockholm University, SE-106 91 Stockholm, Sweden\\ 
$^{5}$Institut f\"ur Astro- und Teilchenphysik and Institut f\"ur Theoretische Physik, Leopold-Franzens-Universit\"at\\
 Innsbruck, A-6020 Innsbruck, Austria\\ 
$^{6}$W. W. Hansen Experimental Physics Laboratory, Kavli Institute for Particle Astrophysics and Cosmology, \\
 Department of Physics and SLAC National Accelerator Laboratory, Stanford University, Stanford, CA 94305, USA \\
$^{7}$Istituto Nazionale di Fisica Nucleare, Sezione di Trieste, I-34127 Trieste, Italy \\
$^{8}$Dipartimento di Fisica, Universit\`a di Trieste, I-34127 Trieste, Italy \\
$^{9}$Exzellenzcluster Universe, Technische Universit\"at M\"unchen, D-85748 Garching, Germany 
}

\begin{document}

\date{Accepted... Received...; in original form \today}

\pagerange{\pageref{firstpage}--\pageref{lastpage}} \pubyear{2010}

\maketitle

\label{firstpage}

\begin{abstract}


The emission from a gamma-ray burst (GRB) photosphere can give rise to a variety of spectral shapes. The spectrum can retain the shape of a Planck function or it can be broadened and have the shape of a Band function.
This fact is best illustrated by studying GRB090902B: The main $\gamma$-ray spectral component is initially close to a Planck function, which can only be explained by emission from the jet photosphere. Later, the same component evolves into a broader Band function.   This burst thus provides  observational evidence that the photosphere can give rise to a non-thermal spectrum. 
We show that such a broadening is most naturally explained by subphotospheric dissipation in the jet. The broadening mainly depends on the strength and location of the dissipation,  on the magnetic field strength, and on the relation between the energy densities of thermal photons and  of the electrons. We suggest that the evolution in spectral shape observed in GRB090902B is due to a decrease of the bulk Lorentz factor of the flow, leading to the main dissipation becoming subphotospheric.  Such a change in the flow parameters  can also explain the correlation observed between the peak energy of the spectrum and low-energy power law slope, $\alpha$, a correlation  commonly observed in GRBs. We conclude that photospheric emission could indeed be a ubiquitous feature during the prompt phase in GRBs and play a decisive  role in creating the diverse spectral shapes and spectral evolutions that are observed.
\end{abstract}


\section{Introduction}

The original fireball model of gamma-ray bursts (GRBs) predicts a
strong photospheric component during the prompt phase (Goodman 1986,
Paczy\'nski 1986). The very high optical depth to scattering expected
near the base of the flow {implies that, regardless of the exact
nature of the emission process, 
the resulting spectrum thermalises and is observed as} a
Planck spectrum. However, only a few GRBs have been identified to be
dominated by a Planck spectrum (Ryde 2004).
Non-thermal spectra are more typically observed (Preece et al. 1998,
Kaneko et al. 2006).  Moreover, if the photosphere occurs far from
where the acceleration of the flow ceases (the saturation radius), the
thermal component {is} weakened by adiabatic expansion. Most of the
flow energy { is} then in the form of kinetic energy and only a
thermal relic {is} left.
 
It was therefore argued that the dominating emission mechanism should
instead be optically-thin synchrotron-emission (Tavani 1996), {
  emitted by relativistic electrons. These  are accelerated following
  kinetic energy dissipation that takes place above the
  photosphere. Such dissipation could, for instance,  result from internal shocks within
the flow (e.g., Rees \& M\'esz\'aros 1994). However, this paradigm has,
in its turn, several severe problems. Foremost, { in order to
  reproduce the observed sub-MeV spectral peak by synchrotron
  emission, a strong magnetic field, typically of the order $B\sim
  10^5-10^6$~G is required. In such a strong magnetic field}, the shocked
electron population is expected to cool rapidly, {producing a
  typical spectrum } which is in stark contradiction to the observed
spectral shape (Crider et al. 1997, Preece et al. 1998, Ghisellini \&
Celotti (1999); see further discussion in e.g. Ryde et al. 2006).

The challenges for optically-thin synchrotron emission { led} to
the revival of the idea that the jet photosphere {may} play an
important role, in one way or another, in the formation of the
spectrum (Eichler \& Levinson 2000, M\'esz\'aros \& Rees, 2000).
M\'esz\'aros \& Rees (2000) proposed that several spectral components
exist in the gamma-ray band, including a Planck spectrum from the
photosphere. Such a spectrum {can} appear as a Band function when
it is observed in a narrow energy band. Indeed, Ryde (2005), who
studied time-resolved spectra from subpulses, showed that in many
cases {the GRB spectra in the BATSE energy range} ($\sim$ 25 -
1900 keV) are {statistically} indistinguishable between fits with
a Band function and with a Planck function combined with a power-law
(BB+pl). In many cases BB+pl model is even preferred over the Band
function.  In addition, the thermal component was found, in these fits,
to have a recurring behaviour during individual pulses:  
the temperature decay {is well fitted} by a characteristic broken power-law 
in time (see further Ryde \& Pe'er, 2009).
Even though the BATSE observations were limited by the narrow energy
band, they thus gave an indication that a photospheric emission
{does, in fact,} {exist} in many bursts, and that GRBs in general have
several spectral components in the gamma-ray band.

Moreover, Ryde \& Pe'er (2009) argued that GRB spectra 
should typically be more complicated than a single Band function, when observed
over a broader energy range. This is because approximately 10\% of all
CGRO BATSE bursts have, during their entire duration, time-resolved
spectra with a Band function {high-energy spectral-slope} $\beta > -2$ (Kaneko et
al. 2006). 
{The peak in the energy flux ($\nu F_\nu$) of these bursts} must therefore be above the BATSE energy
range, at an energy higher than the determined spectral break.  Ryde
\& Pe'er (2009) also studied the few bursts for which there are
simultaneous and time-resolved data available from both BATSE and
EGRET-TASC. They found that several breaks indeed exist in the
spectrum (see also Barat et al. 1998). While the overall power peak,
in the studied bursts, lay in the EGRET range, Ryde \& Pe'er (2009)
interpreted the break in the BATSE range as a subdominant thermal
peak.  {Recently,} similar conclusions were drawn by Guiriec et
al. (2011) and Zhang et al. (2010).

There are thus strong arguments, both theoretical and observational,
that the photospheric emission plays an important role in the spectral
formation during the prompt phase in GRBs. Indeed, recently the Fermi
Gamma-ray Space Telescope (energy range 8 keV $-  > 300$ GeV) has observed
bursts in which a photospheric component is present and several
spectral components are required (Ryde et al. 2010, Guiriec et
al. 2011). However, many bursts observed by the Fermi instruments do not
show such obvious distinction between a Planck function and other
spectral components. The spectral peak is more typically described by a
more broadly peaked Band component.

This has, in its turn, led to increased interest in the suggestion put
forward by Rees \& M\'esz\'aros (2005), who argued that strong
dissipation should naturally occur below and close to the photosphere
(see further \S \ref{sec:3}). { This results from oblique shock
  waves, that are formed at the edges of the jet as it propagates
  through the star. The dissipated energy results in} reprocessing of
the original Planck spectrum, { due to, e.g., Comptonization by
  energetic electrons  (Pe'er et al. 2005, 2006; see also Giannios 2006). 
  This alters the Planck spectrum into
the observed spectrum. { Emission from
  the photosphere} could therefore have a Band-like character and its
shape should depend on the details of the dissipation of the kinetic
energy of the flow. Similar scenarios have recently been
  discussed by several authors, e.g., Beloborodov
et al. (2010), Ioka (2010), Lazzati \& Begelman (2010),  Toma et al. (2010), and Bromberg et
al. (2011); see further discussions in  Ramirez-Ruiz (2005) and Ruffini et al. (2005).

In this paper, we study the particular spectral evolution of 
GRB090902B. In this burst,  the main spectral component, stemming from the photosphere, 
exhibits a change in spectral characteristics  half-way through the prompt phase.
At early times, the thermal component resembles a Planck function, while at late times 
this component broadens significantly. 
Based on this study, we further discuss the conditions under which
photospheric emission is broadened. We argue that this mechanism can be 
applicable to more typical spectral evolutions, in which the spectra 
more gradually evolve from being hard to becoming softer.
In particular,  we argue that it may provide a natural explanation for
the observed variety of spectral shapes in GRBs (in particular the
width of the spectral peak).  In \S 2 the spectral behavior of
GRB090902B is presented and in \S 3 the effects of subphotospheric
shocks on GRB spectra are discussed. In \S 4 we use subphotospheric
shocks to explain the spectral evolution in GRB090902B. We discuss our results in \S 5,
and we conclude in \S 6.

\begin{figure}
\begin{center}
\rotatebox{0}{\resizebox{!}{60mm}{\includegraphics{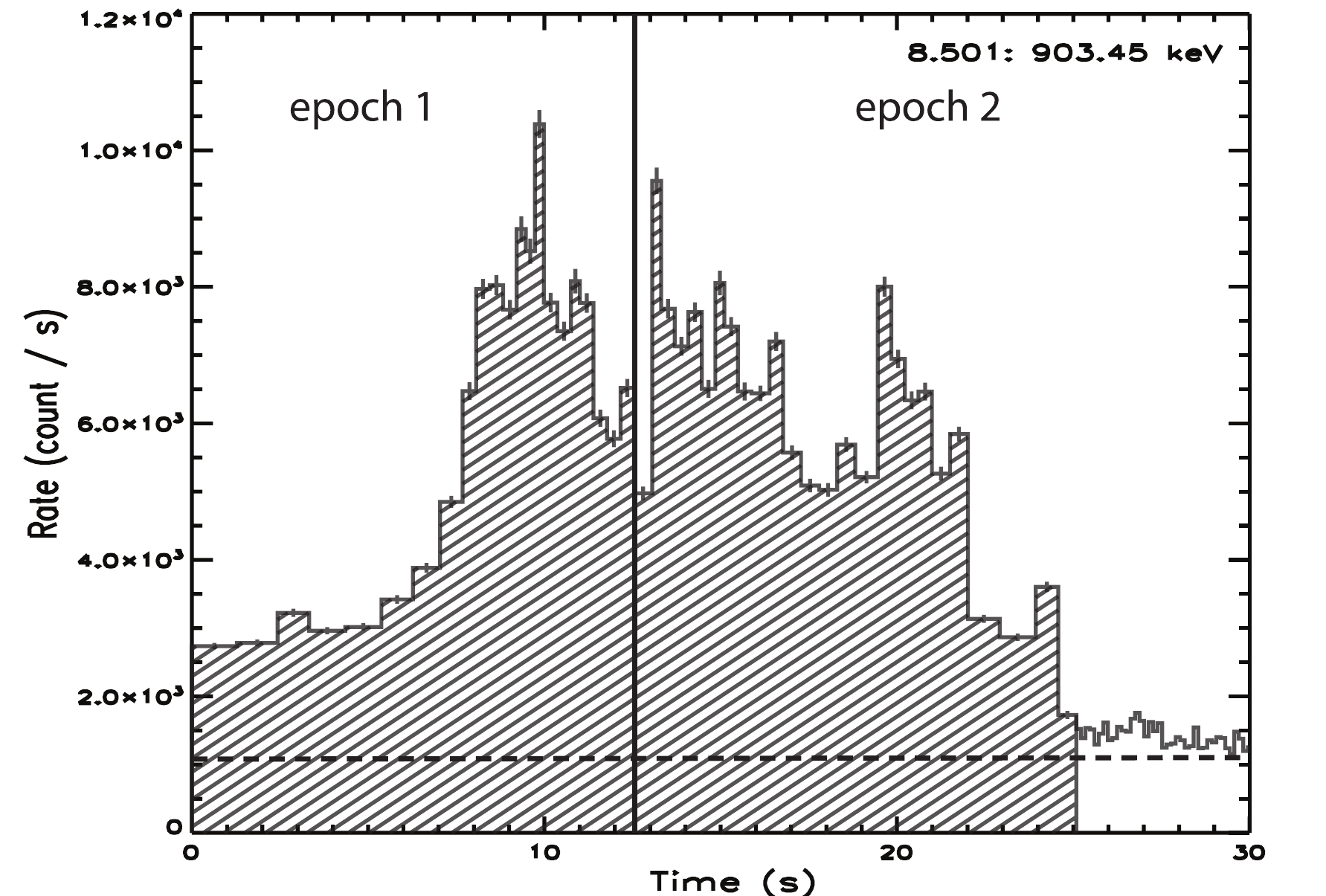}}}
\caption{\small{Light curve of GRB090902B, observed by NaI 1 in the energy range 8.5-904 keV. The presented time-binning is used in the analysis and the two epochs discussed in the paper are {divided} by the vertical line. The thin dashed line is the background level.}}
\label{fig:lc}
\end{center}
\end{figure}

\section{Spectral Behavior of GRB090902B}

The bright and  long burst GRB090902B was detected by the Fermi Gamma-ray  Space Telescope by its two instruments - the Large Area Telescope (LAT; energy range 100 MeV  $-  > 300$ GeV) and the Gamma-ray Burst Monitor (GBM; 8 keV - 40 MeV). The burst lies at a redshift of $z=1.822$ (Cucchiara et al. 2009). It is one of the strongest bursts detected by Fermi and the emission at energies larger than 8 keV lasted for approximately 25 s. The most energetic photon with an energy of $33.4^{+2.7} _{-3.5}$ GeV was detected at 82 s after the trigger by the LAT. The light curve of the prompt phase is shown in Figure \ref{fig:lc}.

During the prompt phase of approximately 25 s two distinct, separate components are observed throughout the duration: a peaked MeV component (modeled with a Band function) and a power-law component. The power-law component is clearly detected at energies both below and above the MeV peak (observations are made in the range of 8 keV - $\sim $ 30 GeV). Moreover, while the MeV peak undergoes substantial spectral evolution the power-law component remains relatively steady with photon-index of approximately -1.9 (Abdo et al. 2009).

At early times (first 12.5 s; epoch 1) the  pronounced MeV peak is so steep and narrow, that it must be attributed to emission from the photosphere (e.g. Fig. 1 in Ryde et al. 2010, see also \S \ref{sec:alpha0}).   However, during the second half of the prompt phase, which lasted for another 12 s (epoch 2), the MeV peak differs significantly from a Planck function, resembling a typical Band spectrum\footnote{Note that a Planck spectrum can in principle be approximated by a ``Band''  function, but with very steep $\alpha=1$ and $\beta \rightarrow - \infty$. We use the
  term ``Band'' to describe spectrum that is not as steep.}.
    

{ We argue here that, although the spectral shape of the MeV peak varies, it is most likely that it has the same origin throughout the burst. The reasons are the following. (i) The spectra are clearly separated into two spectral components throughout the burst duration, namely an MeV peak and an independent power-law component (Fig. \ref{fig:2spectra}). (ii) Although the MeV peak broadens, the power-law component remains relatively steady (see Abdo et al. 2009). (iii) The spectra of the MeV bump during epoch 2 are still inconsistent with the expected non-thermal spectrum; fast cooling electrons yields $\alpha = -1.5$ (see \S \ref{sec:alpha0} below). (iv)  To get a synchrotron peak-energy to lie in a similar energy range as the thermal peak during epoch 1 requires an  unreasonable coincidence. (v) On the other hand, as we show in  \S \ref{sec:4} below, broadening of the thermal peak by sub-photospheric dissipation easily reproduces the observed spectrum. 
 Therefore, we argue that  this burst  provides an observational evidence that the emission from a GRB photosphere does 
not necessarily need to be a narrow, Planck-like spectrum but can be significantly broadened.
This burst is the best example available to study the details of the photosphere and its emission.

In general, the spectral evolution in GRBs is more gradual than observed in GRB090902B. The steepest sub-peak slopes (largest values of $\alpha$) are typically found  at the very beginning of the prompt phase, and only during a small fraction of the burst duration.  The spectra thereafter rapidly soften (Crider et al. 1997,  Ghirlanda et al. 2003).  The particular property of the spectral evolution in GRB090902B is the substantial fraction of the burst duration during which the emission spectrum is Planck-like. This allows  the establishment  of its photospheric origin. }

Here, we further study the time-resolved spectra by following the analysis
performed in Ryde et al. (2010). We use the same detectors (NaI 0, 1
and BGO 0,1 and LAT front and back) and the time binning was
determined by requiring a signal-to-noise ratio in the strongest
illuminated detector to be at least SNR = 45. The data were fit using
RMFIT\footnote{R. S. Mallozzi, R. D. Preece, \& M. S. Briggs, "RMFIT, A Lightcurve and Spectral Analysis Tool,"  \copyright  Robert D. Preece, University of Alabama in Huntsville.} version 3.0 using the Castor C-statistic to determine the 
goodness-of-fit.  The light curve with the time binning used is shown
in Figure \ref{fig:lc}. 
Here the burst is divided into two epochs: one comprising data from within 12.5 s of the trigger, referred to as epoch 1 (analyzed in Ryde et al. 2010), and another comprising data after 12.5 s until the end of the burst, referred to as epoch 2.

\begin{figure*}
\begin{center}
\resizebox{84mm}{!}{\includegraphics{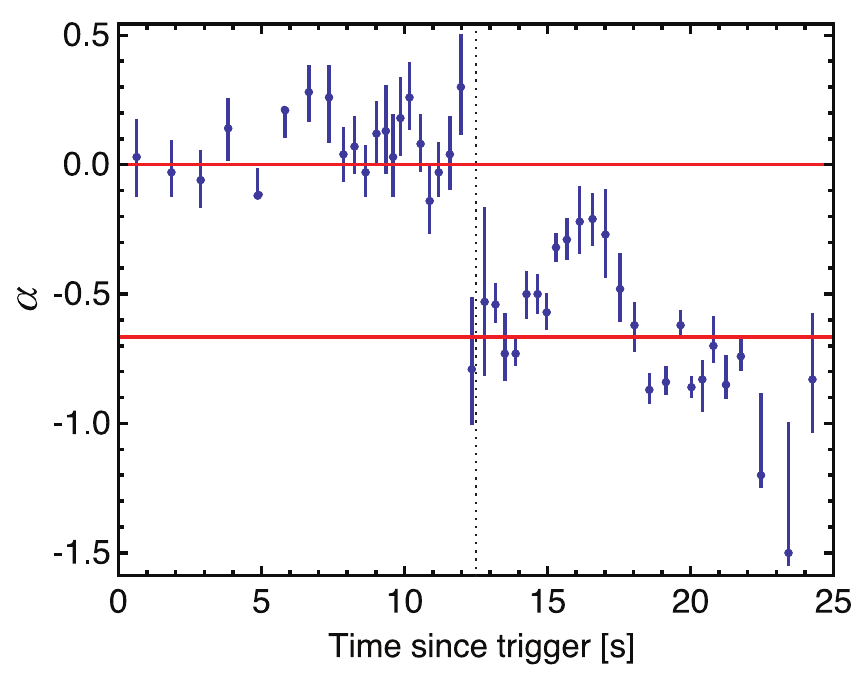}}
\resizebox{84mm}{!}{\includegraphics{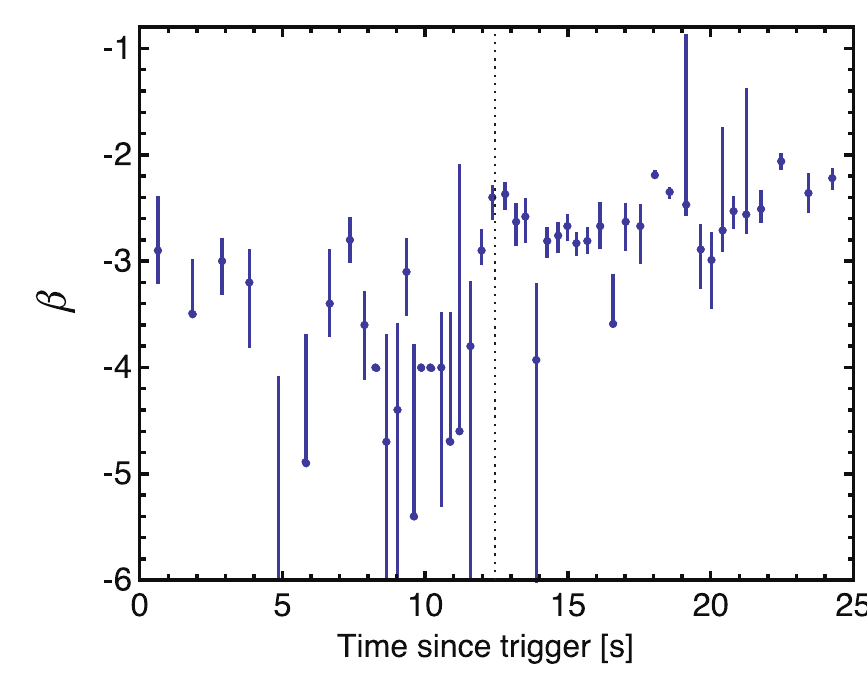}}
\resizebox{87mm}{!}{\includegraphics{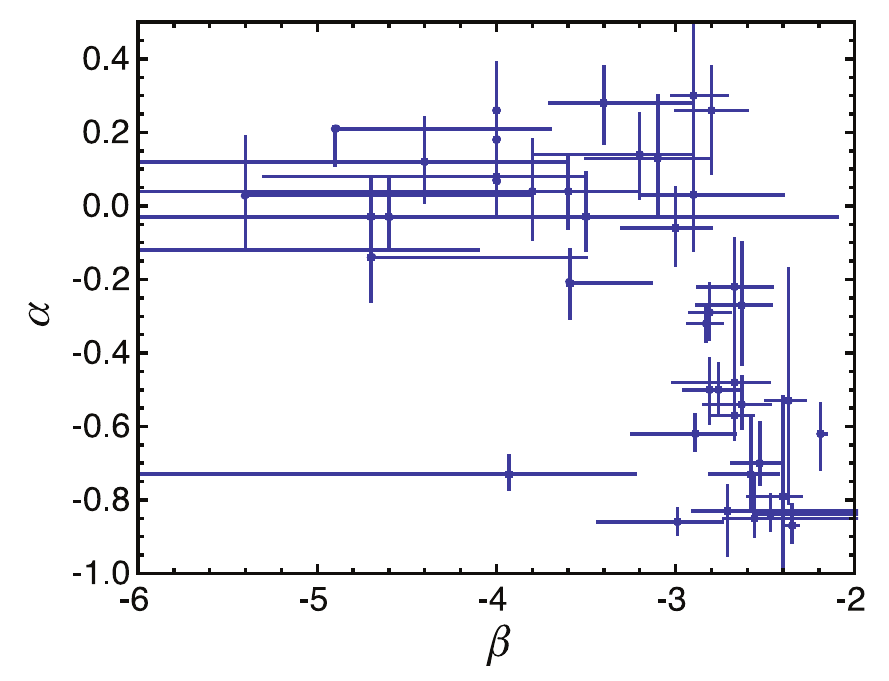}}
\resizebox{84mm}{!}{\includegraphics{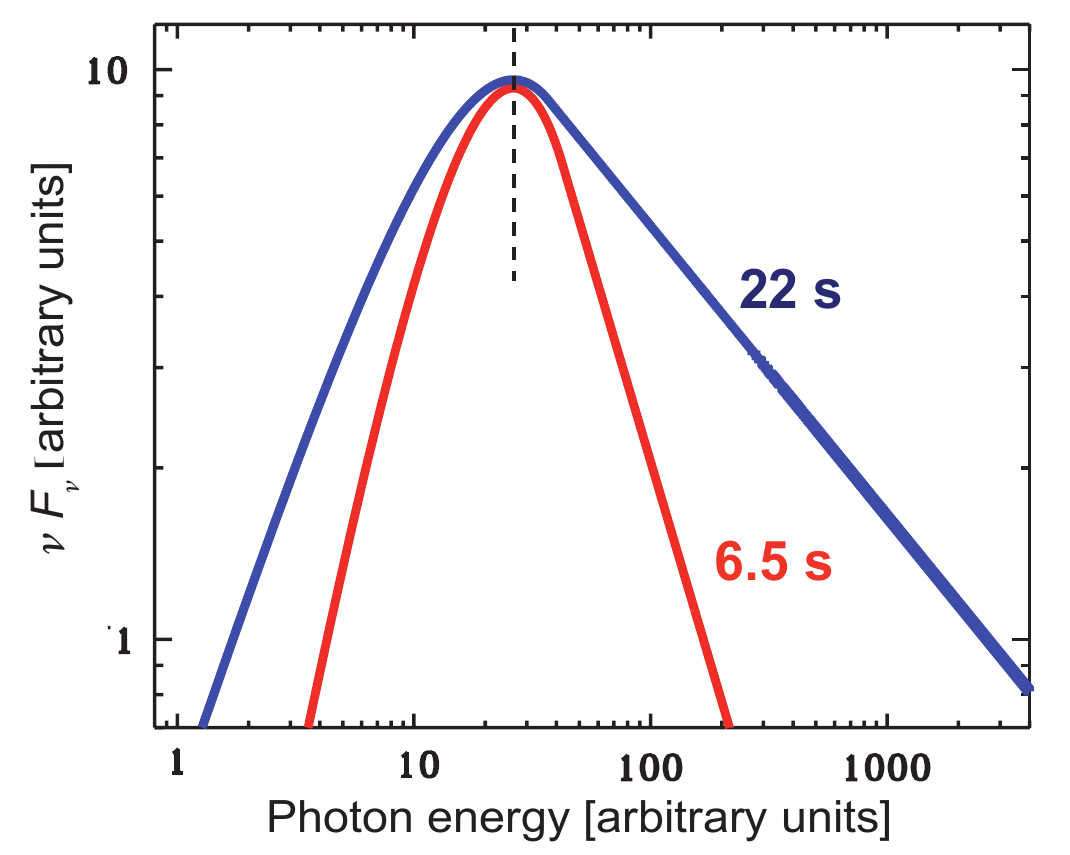}}
\caption{\small{Evolution of the MeV component in GRB090902B. {\it Upper left panel.} Evolution of low-energy photon-index $\alpha$. Two horisontal lines are shown, which correspond to $\alpha = 0$, the most extreme value expected for inverse Compton models and $\alpha = -2/3$, expected for optically-thin synchrotron emission for a slow cooling electron population. The dashed line indicates 12.5 s. {\it Upper right panel.}  Evolution of the high-energy power-law index $\beta$. {\it Lower left panel.} Correlation between $\alpha$ and $\beta$. Note that some points have only  one-sided error bars indicating that they are unconstrained in the other direction.  {\it Lower right panel.}  Peak-aligned Band functions corresponding to the fits at two different times, illustrating the spectral broadening. These two  times include  the narrowest and the broadest Band spectra.}}
\label{fig:1}
\end{center}
\end{figure*}

\subsection{Band function fits}
\label{sec:bandfunc}

A model consisting of  a {Band function (Band et al. 1993)} (for the dominating MeV component) and a power-law component fits the time-resolved spectra well during epoch 1 (Ryde et al. 2010). This is also the case for our fits. However, inspection of the C-stat maps of the parameters  reveals that the error ranges are not well constrained in several bins. The reason for this was determined  to be the low amplitude of the power-law in these bins. In these cases, we froze the power-law amplitude to  the value found in the fit, and then performed a new fit to determine values and uncertainty intervals for the remaining parameters. For all bins, the values of the parameters found before freezing the power-law amplitude were consistent with the values found after the amplitude was frozen.

Table 1 shows the results of our fits to the time-resolved spectra indicating $\alpha$ (photon-index of the sub-peak power-law), $\beta$ (photon-index of the super-peak power-law), $E_{\rm p}$ (peak energy), and the photon-index of the power-law component. 
Figure \ref{fig:1} shows the evolution of the shape of the MeV component. The upper panels show the parameters $\alpha$ and $\beta$  over the full duration of the burst (0-25 s). The index of $\alpha = -2/3$ expected from optically-thin synchrotron emission is indicated  (Rybicki \& Lightman 1979) as well as $\alpha = 0$, the spectral slope expected from, e.g.  jitter radiation and from extreme cases of inverse Compton emission from a delta function distribution of electrons (Jones 1968). The errors on the data points represent asymmetric, one-sigma uncertainties on the parameter values found from the fitting.  

In Figure \ref{fig:1}, it is clearly seen that the spectral shape exhibits a  change in character  at
approximately 12.5 s after the trigger, as noted already by Ryde
et al. (2010). From being very peaked, with $\alpha \sim 0.3$ and
$\beta \sim -3.5$ and with a {spectral width} of the peak\footnote{We here define the spectral width, $w$, as the ratio $E_{\rm high}/E_{\rm low}$,  where energy fluxes $F_E(E_{\rm low}$) and $F_E(E_{\rm high}$) are equal to half the peak flux ($F_{\rm peak}/2$) below and beyond the peak, respectively.} $w$
=  6, the spectrum broadens significantly. {During
  the second epoch, the} typical value of $\alpha$ $\sim -0.6$ and
$\beta \sim -2.5$.  {As a result, the spectral width has increased between the
  first and second epochs} by more than a factor of two, to typically
$w = $10 -- $20$.  In the lower right-hand panel in Figure \ref{fig:1} the best-fit
Band function of the MeV peak is plotted for two instants, at
6.5 s and at 22 s after the GBM trigger. These spectra have been aligned 
to each other's $E_{\rm p}$ values in order to highlight the spectral broadening: 
while the {typical spectrum at} epoch 1 is
close to a Planck function, the {spectrum at} epoch 2 has a shape that is typical
for a GRB spectrum, that is, a Band function.
 %

{Even though the change in spectral character between the epochs is clear from the above discussion, there are still important similarities between the epochs.  Most notably, Abdo et al. (2009)  clearly showed that  the power-law component remains relatively steady during the spectral evolution of the prompt phase. Moreover, during the time period 15 -- 17 s, the MeV component becomes relatively hard again.  The similarities between these spectra and the epoch 1 spectra are shown in Figure \ref{fig:2spectra}. Here a spectrum from epoch 1 (8.1-- 8.5 s) and a spectrum from epoch 2 (15.9 -- 16.4 s) are shown. The broadening of the MeV component, compared to the epoch 1 spectra, is still apparent, even though the broadening is not as large as for the other  epoch 2 spectra. These observations are thus strong indications of that the emission during the two epochs are of similar origin, i.e. a photospheric  and an optically-thin component. }

\begin{figure*}
\begin{center}
\resizebox{84mm}{!}{\includegraphics{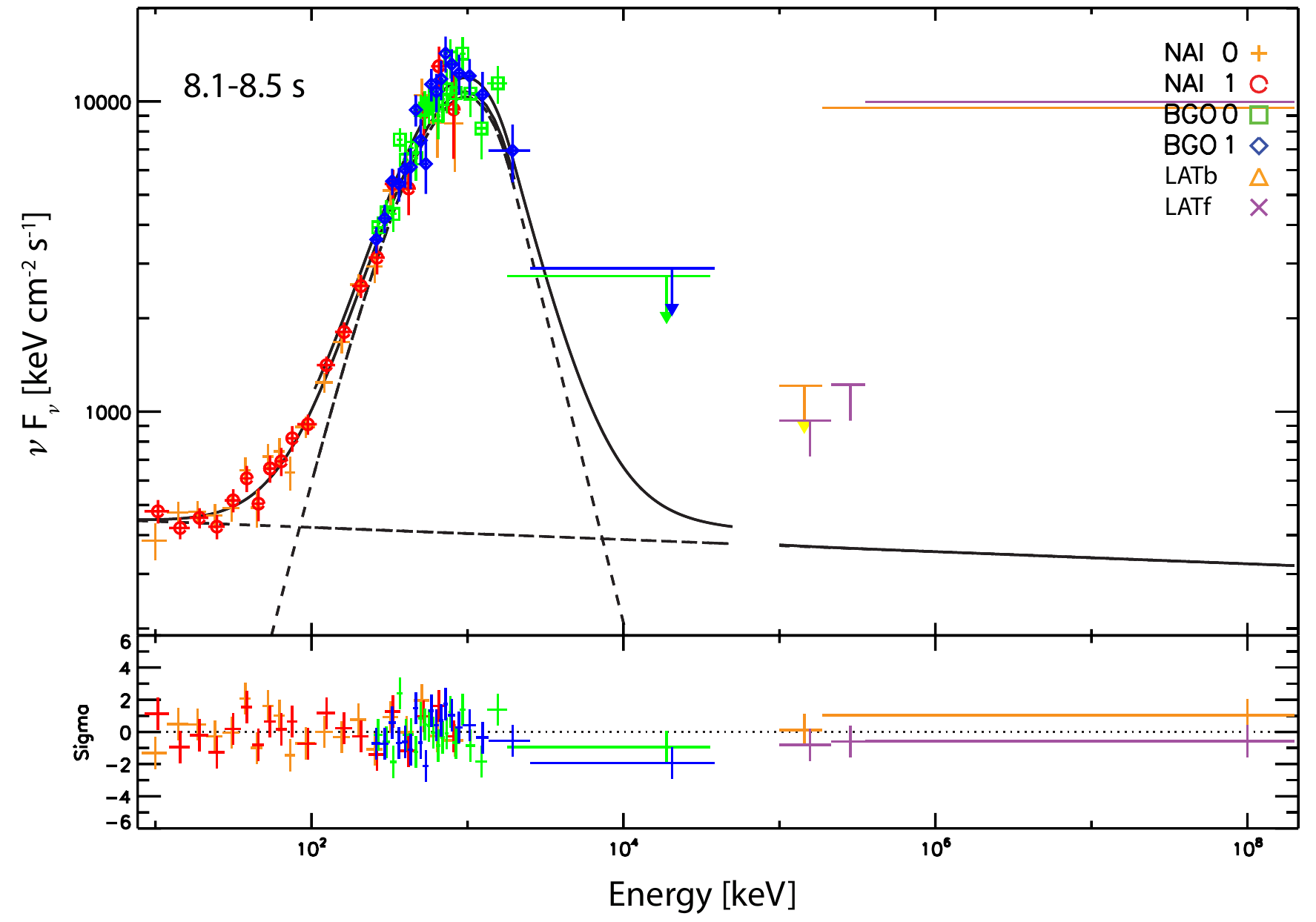}}
\resizebox{84mm}{!}{\includegraphics{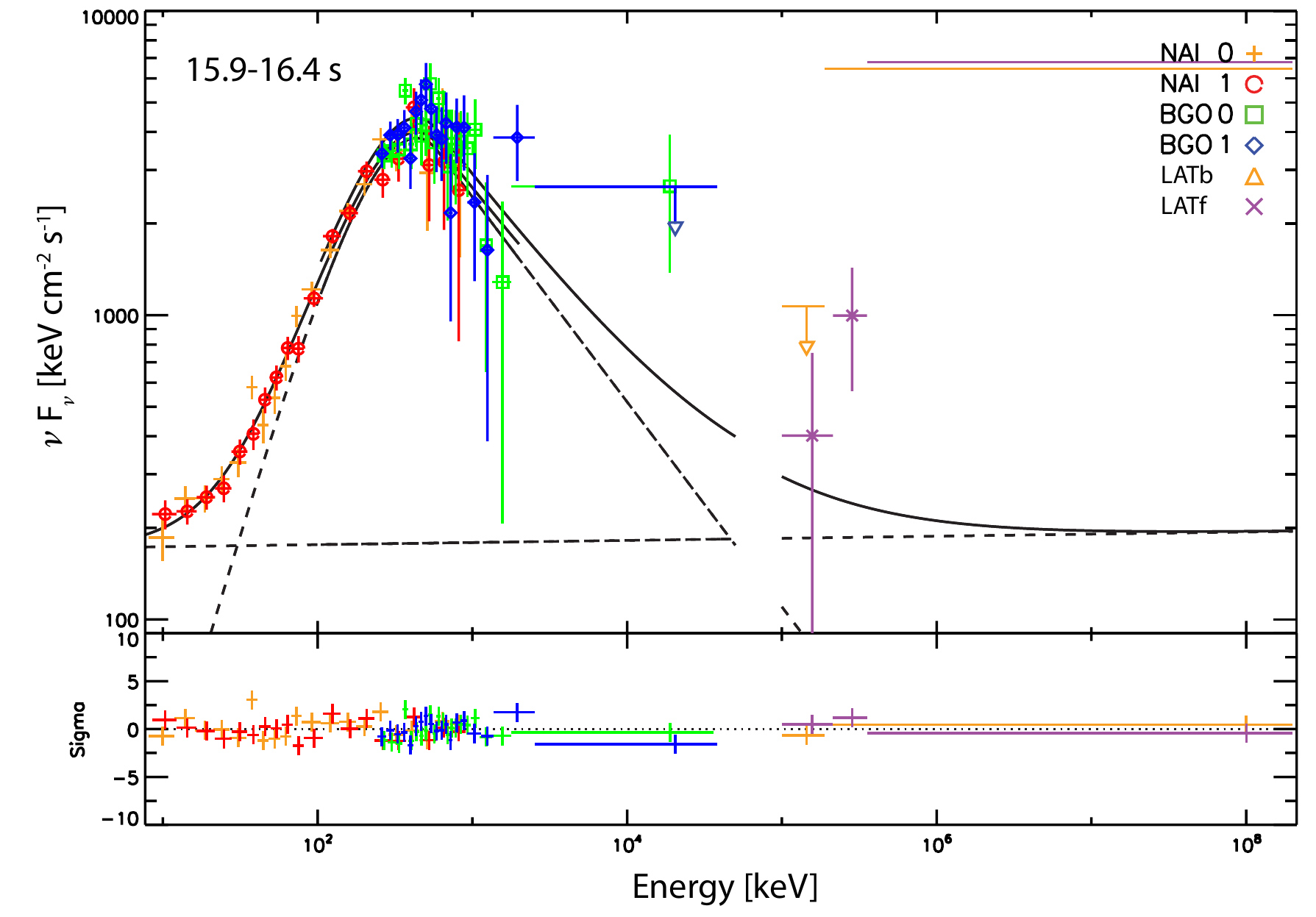}}
\caption{\small{Time resolved $\nu F_{\nu}$ spectrum for two time intervals $t = 8.1- 8.5$ s (epoch 1) and  $t= 15.9-16.4$ s (epoch 2).  The Band  + power-law model is fitted to the data over the GBM + LAT 
energy ranges. The symbols refer to the different instruments on the  Fermi Gamma-ray Space Telescope. While there still are  similarities between the spectra, the broadening of the MeV component is apparent  (Compare to  Fig. 1 in Ryde et al. 2010).
}}
\label{fig:2spectra}
\end{center}
\end{figure*}

Figure \ref{fig:4} shows the $E_{\rm p}$ evolution as a function of
time. The averaged value of  $E_{\rm p}$ is lower during epoch 2
compared to epoch 1. 


In spite of the variations in the spectrum of the MeV component,  it dominates the spectral energy flux throughout the burst duration.
The ratio of the energy flux in the MeV peak relative to the total Fermi $\gamma$-ray flux  (MeV peak + power law component) is in the range 80 and 95 \% during the entire burst duration. This mainly results from a decrease of the amplitude of the PL component towards the end.

\begin{figure}
\begin{center}
\rotatebox{0}{\resizebox{!}{60mm}{\includegraphics{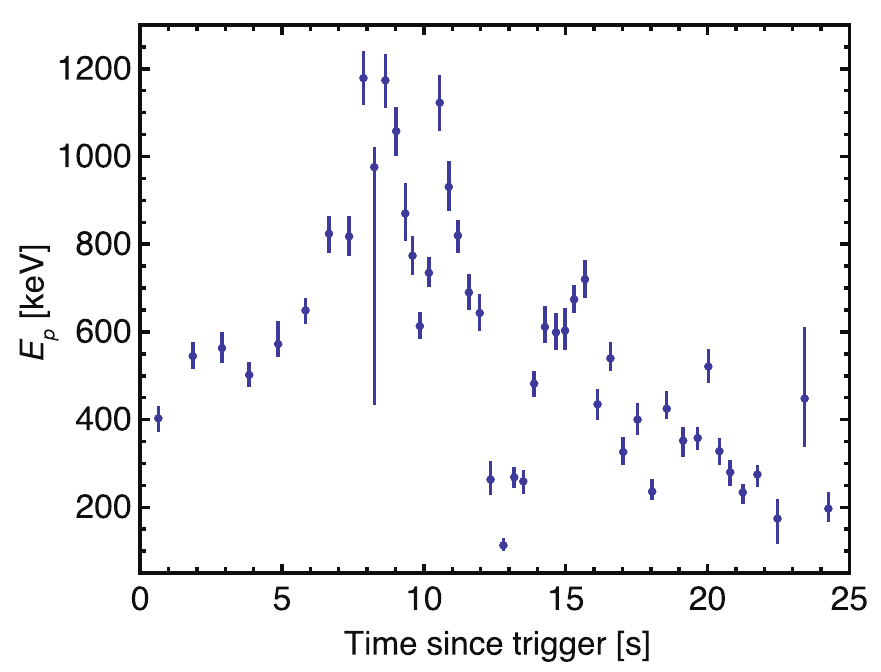}}}
\caption{\small{The evolution of the peak energy of the Band function fits, $E_{\rm p}$, as a function of time since the GBM trigger.}}
\label{fig:4}
\end{center}
\end{figure}

\begin{table*}
\begin{tabular}{l l l l l l} 

\hline
\hline

Time (s) & PL index & $\rm E_{peak}$ & $\alpha$ & $\beta$ & C-stat/dof \\

\hline

12.54-13.06 & $-2.04_{-0.07}^{+0.09}$ & $113_{-10}^{+13}$  & $-0.53_{-0.28}^{+0.36}$ & $-2.37_{-0.13}^{+0.10}$ & 589.69/598 \\

13.06-13.31 & $-1.65_{-0.51}^{+0.12}$ & $268_{-21}^{+21}$ & $-0.54_{-0.07}^{+0.08}$ & $-2.63_{-0.22}^{+0.16}$ & 548.28/598 \\

13.31-13.70 & $-1.97_{-0.13}^{+0.16}$ & $259_{-24}^{+22}$ & $-0.73_{-0.10}^{+0.15}$ & $-2.58_{-0.23}^{+0.16}$ & 524.54/598 \\

13.70-14.08 & $-1.78_{-0.09}^{+0.08}$ & $482_{-26}^{+24}$ & $-0.73_{-0.04}^{+0.05}$ & $-3.93_{-17.6}^{+0.71}$ & 517.21/598 \\

14.08-14.46 & $-2.88_{-0.83}^{+0.43}$ & $611_{-33}^{+44}$ & $-0.50_{-0.10}^{+0.09}$ & $-2.81_{-0.15}^{+0.12}$ & 551.20/597 \\

14.46-14.85 & $-2.46_{-0.04}^{+0.04}$ & $599_{-37}^{+41}$ & $-0.50_{-0.07}^{+0.08}$ & $-2.76_{-0.15}^{+0.12}$ & 585.00/598 \\

14.85-15.10 & $-3.87_{-0.98}^{+2.82}$ & $603_{-40}^{+48}$ & $-0.57_{-0.07}^{+0.07}$ & $-2.67_{-0.13}^{+0.11}$ & 445.00/598 \\

15.10-15.49 & $-3.12_{-0.04}^{+0.04}$ & $674_{-28}^{+29}$ & $-0.32_{-0.05}^{+0.05}$ & $-2.83_{-0.11}^{+0.10}$ & 617.42/598 \\

15.49-15.87 & $-1.94_{-0.03}^{+0.04}$ & $720_{-38}^{+41}$ & $-0.29_{-0.08}^{+0.08}$ & $-2.81_{-0.12}^{+0.12}$ & 584.36/598 \\

15.87-16.38 & $-1.99_{-0.73}^{+0.06}$ & $435_{-31}^{+30}$ & $-0.22_{-0.12}^{+0.13}$ & $-2.67_{-0.21}^{+0.22}$ & 559.96/597 \\

16.38-16.77 & $-1.88_{-0.05}^{+0.05}$ & $540_{-26}^{+33}$ & $-0.21_{-0.10}^{+0.10}$ & $-3.59_{unc}^{+0.46}$ & 608.65/597 \\

16.77-17.28 & $-1.88_{-0.06}^{+0.05}$ & $326_{-25}^{+30}$ & $-0.27_{-0.16}^{+0.17}$ & $-2.63_{-0.26}^{+0.17}$ & 613.26/597 \\

17.28-17.79 & $-1.99_{-1.21}^{+0.11}$ & $400_{-31}^{+34}$ & $-0.48_{-0.12}^{+0.13}$ & $-2.67_{-0.35}^{+0.20}$ & 539.27/597 \\

17.79-18.30 & $-5.90_{-2.08}^{+1.41}$ & $236_{-16}^{+24}$ & $-0.62_{-0.10}^{+0.09}$ & $-2.19_{-0.02}^{+0.04}$ & 574.43/597 \\

18.30-18.82 & $-5.40_{unc}^{unc}$ & $425_{-20}^{+37}$  & $-0.87_{-0.05}^{+0.06}$ & $-2.35_{-0.06}^{+0.04}$ & 536.23/597 \\

18.82-19.46 & $-1.38_{-0.16}^{+0.09}$ & $352_{-34}^{+26}$ & $-0.84_{-0.05}^{+0.06}$ & $-2.47_{-0.10}^{+1.59}$ & 723.50/598 \\

19.46-19.84 & $-1.34_{-0.12}^{+0.09}$ & $358_{-23}^{+22}$  & $-0.62_{-0.05}^{+0.06}$ & $-2.89_{-0.36}^{+0.23}$ & 539.89/598 \\

19.84-20.22 & $-1.41_{-0.12}^{+0.09}$ & $521_{-33}^{+35}$ & $-0.86_{-0.04}^{+0.04}$ & $-2.99_{-0.45}^{+0.26}$ & 488.24/598 \\

20.22-20.61 & $-4.22_{-0.16}^{+0.71}$ & $328_{-28}^{+26}$ & $-0.83_{-0.12}^{+0.07}$ & $-2.71_{-0.20}^{+0.96}$ & 691.51/598 \\

20.61-20.99 & $-1.96_{-0.31}^{+0.27}$ & $280_{-27}^{+23}$ & $-0.70_{-0.06}^{+0.11}$ & $-2.53_{-0.16}^{+0.13}$ & 531.45/598 \\

20.99-21.50 & $-2.07_{unc}^{unc}$ & $234_{-22}^{+15}$ & $-0.85_{-0.05}^{+0.11}$ & $-2.56_{-0.17}^{+1.17}$ & 736.42/598 \\

21.50-22.02 & $-1.70_{-0.29}^{+0.12}$ & $275_{-25}^{+18}$ & $-0.74_{-0.05}^{+0.07}$ & $-2.51_{-0.12}^{+0.17}$ & 756.01/598 \\

22.02-22.91 & $-2.15_{unc}^{+0.81}$ & $ 174_{-53}^{+41}$ & $-1.25_{-0.05}^{+0.31}$ & $-2.06_{-0.07}^{+0.06}$ & 639.88/598 \\

22.91-23.94 & $-1.39_{-0.10}^{+0.07}$ & $ 448_{-107}^{+160}$ & $-1.50_{-0.05}^{+0.05}$ & $-2.36_{-0.17}^{+0.17}$ & 807.17/598 \\

23.94-24.58 & $-2.13_{-0.12}^{+0.27}$ & $ 197_{-27}^{+33}$ & $-0.82_{-0.20}^{+0.25}$ & $-2.22_{-0.10}^{+0.08}$ & 546.56/598 \\

\hline
\end{tabular}
\label{table:fits}
\caption{Results of spectral fits to the data of GRB090902B during the time 12.54-24.58 s after the burst trigger (epoch 2). Uncertainties marked {\it unc} indicate that
the parameter is unconstrained. Results for epoch 1 are presented in Ryde et al. (2010).}
\end{table*}

\subsection{Significance of the hard $\alpha$-values.}
\label{sec:alpha0}

During epoch 1 the averaged value of $\alpha$ is $<\alpha> = 0.11$ and for several
of the time-bins the value of $\alpha$ is even steeper, with $\alpha
\gta 0.2$. These very hard spectra are particular challenging for
non-thermal models. Synchrotron and inverse Compton emission in 
the fast cooling regime are expected to produce a spectral slope of $\alpha = -1.5$.
Here, we consider  the most extreme scenario with $\alpha = 0$ (Jones 1968, Epstein
\& Petrosian 1973). We therefore want to estimate the significance of
rejecting $\alpha = 0$ for these spectra.  This allows us to determine the
significance of {the conclusion that neither synchrotron nor inverse Compton scattering 
processes can explain the spectra.} 

This is done by simulating spectra with RMFIT v3.0 using the set of observed spectral parameter values of the Band function fit.  We use the Fermi detector responses for GRB090902B and take into account  the Poissonian nature of the observed counts and realistic background emission. The simulated count distributions  are then fitted in the same way as the real data. The parameter values we find can then be studied.

We illustrate the procedure on the 7th time-bin (6.3-7.0 s) which has 
the hardest observed value of  $\alpha= 0.3 \pm 0.1$ (Ryde
et al. 2010).  We start with the null hypothesis that the spectrum observed in this time bin
has an actual value $\alpha = 0$.  We therefore freeze $\alpha$ at this
value, and, based on the best fit parameters we find from the data, we
perform 100,000 simulations.  This large number of simulations allows
us to make a proper estimation of the significance level.  The
simulated spectra are then fitted with the Band function with all
parameters free to vary. The distribution of the {values of $\alpha$}  that
we find are shown in Figure \ref{fig:alphalimit}. The distribution is
approximately Gaussian and is slightly skewed. It has a mean value of
$\alpha = 0.01$ and a standard deviation of 0.068. The inset in Figure
\ref{fig:alphalimit} is a magnification of the distribution around
$\alpha=0.3$. This figure shows that 8 out of the 100,000 simulated
spectra have fits with {values of $\alpha$} greater than 0.3. This gives
the probability that a spectrum with an actual value $\alpha=0$ should
be observed with $\alpha > 0.3$ by chance to be $8 \times
10^{-5}$. Therefore the null hypothesis can be rejected on a very high
confidence level for this bin.  Most of the measured {values of $\alpha$}
during epoch 1 are, though, not significantly inconsistent with
$\alpha =0$ (see Fig. \ref{fig:1}). In any case, the epoch 1 spectra
are challenging for purely non-thermal emission models, since $\alpha = 0$ is only expected under somewhat  extreme conditions, see further \S \ref{sec:51}.  The expected value is rather $\alpha = -1.5$,  which is produced by  a population of fast cooling electrons (Ghisellini \& Celotti, 1999). 

The averaged value of $\alpha$ during epoch 2 is $<\alpha> = -0.65$. This value is still significantly inconsistent with $\alpha = -1.5$.  For instance, the 42nd time-bin (19.84-20.22 s) has one of the softest values of $\alpha$ with $\alpha = -0.86 \pm 0.04$. Even though, when we perform a fit with $\alpha$ frozen at -1.5 to  this time-bin, we find  an increase of the  C-stat value from 488 to 1167 for 597 degrees of freedom. The hypothesis of having $\alpha = -1.5$ can therefore be rejected on a significance level of less than $ 10^{-10}$.  We also point at the fact that several of the fitted spectra in epoch 2 have $\alpha \sim -0.2$; the 32nd time-bin (15.87-16.38 s)  has the hardest spectrum with $\alpha = - 0.2 \pm 0.1 $, approaching the values of $\alpha$ from epoch 1, see Figure 2. 

\begin{figure}
\begin{center}
\resizebox{87mm}{!}{\includegraphics{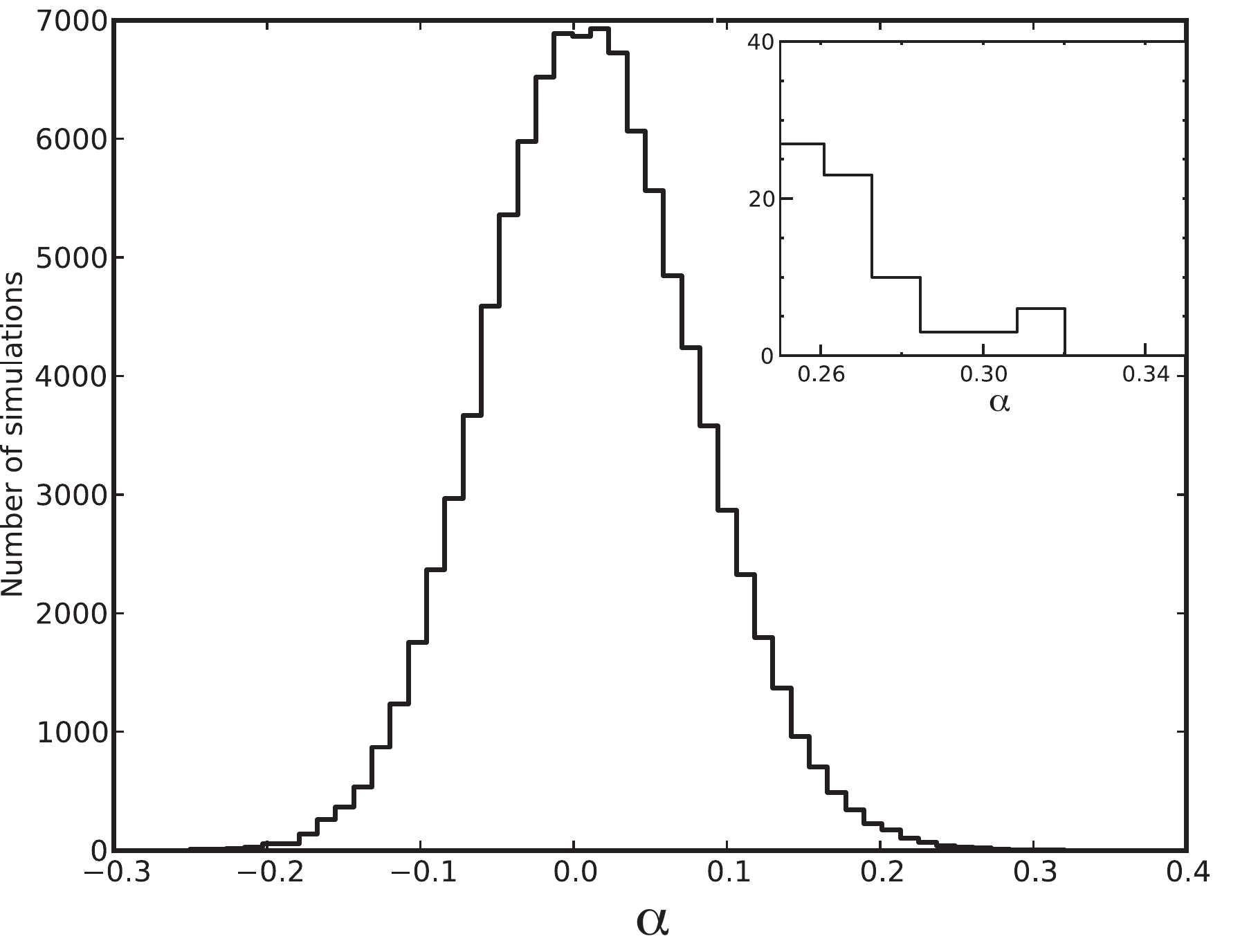}}
\caption{\small{{Histogram distribution of $\alpha$ values  found from 100,000 simulated spectra. The inlay is a magnification of an interval around $\alpha = 0.3$. See the text for details.}}}
\label{fig:alphalimit}
\end{center}
\end{figure}

\section{Subphotospheric Heating and its effect on the emission spectrum}
\label{sec:3}

As we saw above, GRB090902B is a particularly interesting burst since the main spectral component 
initially is close to a Planck function but later evolves into a broader Band function. Indeed, as we will see in this section, the emitted spectrum that is expected from a GRB photosphere depends on the existence of significant dissipation close to the photosphere.

\subsection{Photospheres and dissipation radius}

The photosphere  related to  the electrons associated with the baryons
in the outflow (the baryonic photosphere) is expected to be  at  (e.g. M\'esz\'aros et al. 2002) 
\begin{equation}
r_{\rm ph} \sim 4.8 \times 10^{11}  \frac{L}{10^{53} \mathrm{erg/s}}
\left( \frac{\Gamma}{630} \right)^{-3}  \mathrm{cm},  
\label{eq:ph}
 \end{equation}
where the typical value for GRB090902B is used: $L = 10^{53}$ erg/s
(Abdo et al. 2009). We also use the {\it time-averaged} value for
 $\Gamma =600\, Y^{1/4} \sim 630$ (Ryde et al. 2010)\footnote{
 Later we also use the notation $\Gamma_{2.8} \equiv (\Gamma/630)$ and $L_{53} \equiv (L/10^{53} \rm{erg/s})$:
Note that Pe'er et al. (2010) used $\Gamma \sim 1000$ which was
   estimated for a single time bin used in their analysis}.  Here $Y$
 denotes the ratio between the total fireball energy and the energy
 emitted in $\gamma$-rays and can be estimated from afterglow
 measurements. Cenko et al. (2010) estimated the value to $Y \sim 1.2$
 in the case of GRB090902B.

 Pair formation can be significant and modify the spectrum
 (e.g. Eichler \& Levinson 2000, Pe'er \& Waxman 2004). {
   Subphotospheric} dissipation can lead to {copious production of
   pairs, resulting in a second, pair photosphere above} the baryonic
 photosphere (M\'esz\'aros et al. 2002).  {During the dissipation process, electrons are
  expected to be accelerated to high energies, thereby emitting
   energetic photons at energies above the threshold for pair
   production, $m_e c^2$. These photons then produce pairs by
   annihilating with the lower energy photons. The created pairs have
   modest Lorentz factors, $\gamma_e \sim $ few (in the comoving
   frame).  For sub-photospheric dissipation these pairs are expected
   to be more numerous than the baryon-related electrons, and a pair
   photosphere is expected to be established at radius (Pe'er \& Waxman,
   2004) }
\begin{equation}
r_{\pm}  \sim 8.0  \times 10^{14} \left(  \frac{L}{10^{53} \rm{erg/s}} \right) \epsilon_\pm \, \, \tilde{\alpha}^{-1}  \left( \frac{\Gamma} {630}\right)^{-3}  \mathrm{cm}
\label{eq:phpm}
\end{equation}
where $\epsilon_\pm$ is the fraction of the total fireball energy that
is dissipated into photons with comoving energy larger than $m_e c^2$
and  is available for pair formation.  We
define $r_0 = \tilde{ \alpha} r_{\rm g} $ as the effective radius
at which the outflow starts to accelerate (in the absence of
dissipation): $\tilde{\alpha} \gta 1$ and $r_g\equiv 2GM/c^2 = 3
\times 10^{6} (M/10M_\odot)$~cm is the Schwarzschild radius of the
central black hole of mass $M=10 \, M_\odot$.  The pair photosphere is
above the baryonic photosphere if $\epsilon_\pm > m_{\rm e}/ m_{\rm p}
= 5.45 \times 10^{-4}$ (Rees \& M\'esz\'aros 2005).

{The existence of a dissipation process is in fact required by the data.}
Since we see non-thermal spectra in GRBs this implies that there is
a mechanism that dissipates some fraction of the jet kinetic energy.
The exact nature of this {dissipation} is debatable. {Several models exist in the literature. The leading ones are:  (i)} internal shocks in which shells with varying Lorentz
factors interact with each other (Rees \& M\'esz\'aros 1994); (ii) oblique shocks within the funnel in the star (e.g. Morsony et al.
2007); (iii) collisional dissipation in the flow
(Beloborodov 2003); or (iv) in the case of Poynting-flux dominated flows,
{the dissipation could result from} magnetic reconnection (e.g. Thompson 1994, Giannios \&
Spruit 2005). The shocked region subsequently cools by emitting
photons through, for instance, synchrotron emission and/or inverse
Compton emission.

In the internal shocks {scenario,  dissipative} heating is typically
assumed to occur well above the photosphere: with variations in
Lorentz factor in the flow of the size $\Delta \Gamma \sim \Gamma$ the
internal shocks occur at $r_{\rm sh} \sim 2 r_0 \Gamma^2 \sim 2.1
\times 10^{13}$ cm.
However, Rees \& M\'esz\'aros (2005) pointed out
that due to jet edge effects oblique shocks might form below and close
to the photosphere:
\begin{equation}
r_{\rm sh} \sim 2 r_0 \Gamma ^2 \theta_j \sim 6.3 \times 10^{11}  \,\, \tilde{\alpha} \left( \frac{\Gamma}{630} \right)^2 \frac{\theta_j}{3 \times 10^{-2}} \mathrm{cm}
\label{eq:shock}
\end{equation}
where  $\theta_j$ is the nozzle opening half angle.  

It thus follows that  internal shocks may naturally be expected  in proximity  of the photosphere.  
While the details of these process are still highly uncertain, this problem is extensively being studied numerically. Indeed,
numerical simulations of a jet penetrating though the core of the progenitor show that such shocks do indeed occur (Lazzati et al. 2009, Mizuta et al. 2010). 
We note that similar conclusions are drawn for  other dissipation processes as well (Giannios \& Spruit 2005, Giannios 2006, 2007, Tchekhovskoy et al. 2008, Beloborodov 2010).

\subsection{Broadening of the Planck spectrum: analytical arguments}
\label{sec:Felix}

The energy that is dissipated in subphotospheric shocks {partly
thermalises} again to an extent that { depends on the conditions at
the dissipation site, particularly on the optical depth.}
Detailed calculations of the thermalisation process of such shocks by  Pe'er et al. (2006), and in \S
\ref{sec:asaf} below show that a large variety of spectral shapes can
be achieved. Similar results are found in magnetic reconnection  models (e.g., Giannios 2006).  
In particular, Pe'er et al. (2006) showed that the Planck
function that is injected into the dissipation region is modified to a
varying extent depending on the dissipation fractions and the optical
depth. The Planck spectrum therefore loses its
original shape and the outgoing photospheric emission has a
non-thermal shape. The resulting spectrum can have a rather complex
spectral shape. {As we show here, under plausible conditions in
 many cases it can be described as a smoothly broken power law}.

While detailed numerical results are presented below, we give here
some basic analytical arguments to describe the conditions under which
significant modification of the spectrum can take place.

Assume that the dissipation process, regardless of its exact nature,
produces a population of energetic electrons with characteristic
Lorentz factor $\gamma_e \gg 1$. These electrons cool by Compton
scattering the thermal (photospheric) photons, on a time scale given
by $t_{cool} \simeq m_e c/ (4/3) \gamma_e \sigma_T u_{ph}$. Here,
$u_{ph}$ is the energy density in the thermal photon component, and
$\sigma_T$ is Thomson cross section.  This loss time can be compared
to the dynamical time of the problem, $t_{dyn} = r/\Gamma c$ to obtain
(see Pe'er et al. 2005)
\beq
{t_{loss} \over t_{dyn}} = {3 \over 4} {m_e c^2 \Gamma \over \gamma_e
  \sigma_T u_{ph} r} = {3 \over 4} {u_{el} \over \gamma_e^2 u_{ph}
  \tau_{\gamma e}}.
\label{eq:t_loss}
\eeq 
Here, $r$ is the dissipation radius, $\tau_{\gamma e} = (r/\Gamma) n_e
\sigma_T$ is the optical depth to photon scattering by the electrons
and $u_{el} = \gamma_e n_e m_e c^2$ is the energy density in the
electron component. Equation (\ref{eq:t_loss}) thus implies that,
regardless of the exact nature of the dissipation process, the
electrons cool rapidly, on a time scale much shorter than the
dynamical time. This is provided that (i) the dissipation occurs below or not
too high above the photosphere, and (ii) that the energy that is being
dissipated is not much larger than the energy density in the
photosphere (see Pe'er et al. 2005).  The rapid cooling is due to
Comptonisation of the thermal component; obviously, synchrotron
emission further contributes to the rapid cooling of the electrons.

This rapid cooling implies that the electrons lose most of the energy
imparted to them by the dissipation process. This energy is used to 
up-scatter the photospheric photons, as well as to emit synchrotron 
photons. Part of this energy is converted into
pairs, by upscattered photons that are energetic enough.

The distribution of the rapidly cooled electrons reaches a
quasi-steady state: when the electrons are cold enough, inverse
Compton scattering becomes inefficient, while other processes, such as
direct Compton heating or synchrotron self-absorption heats the (cold)
electrons. The electron distribution can therefore be approximated as
a (quasi-) Maxwellian distribution, with characteristic temperature $T_e$. As long
as the dissipative process that heats the electrons (or introduces a
population of energetic electrons into the plasma) exists, the
steady temperature of the electrons is inevitably higher than the temperature
of the photospheric (thermal) photons: $T_e \gta T_\gamma$.

The plasma is therefore characterized by two temperatures,
$T_{\gamma}$ and $T_e$. Due to the rapid cooling, during most of the
dynamical time the scatterings take place between the thermal photons
and the cold electrons. Since $T_{e} \gta T_{\gamma}$, the thermal
photons gain energy, resulting in modification of the Wien part of the
Planck spectrum, to produce a smoother cutoff at high energies
(above the thermal peak). The exact shape of the spectrum at these
energies (which corresponds to ``$\beta$'' in Band fits), depends on
the optical depth, and the ratio of the energy densities in the
electrons and photon components. A significant shift can be obtained
if the Compton (or equivalently Kompaneet's) $y$-parameter is of the order of a few (which
translates to optical depth of a few), and energy densities in the
electron and photon components are roughly comparable.

Interestingly, somewhat similar conditions are required in order to
obtain a significant modification to the Rayleigh-Jeans part of the
Planck spectrum. This part is modified if two conditions are met:
first, significant number of photons at energies below the thermal
peak must be introduced into the plasma. The most natural way to
obtain a large density of cold photons is via synchrotron radiation.
This emission is expected at low energies, resulting from emission
from the cold electrons. In order to obtain a significant flux, a
strong magnetic field is thus needed. The second condition is that
up-scattering of these photons leads to energies comparable to the
original thermal photons. The condition here is again $y \geq 1$,
which, due to the low value of $T_e$ is translated into $\tau_{\gamma
  e} \gta $ few.

We further note that if the optical depth $\tau_{\gamma e} \rightarrow
\infty$, the spectrum approaches either a Planck or Wien spectrum, as
the energy given to the electrons is distributed among the electrons
and the photons. If the energy given to the electrons by the
dissipation process is much larger than the energy in the thermal
component, $u_{el} \gg u_{ph}$, then the non-thermal part of the
spectrum is significantly more luminous than the thermal part. In such
a scenario, the thermal part may not be detectable (for very high
optical depth, the resulting Planck or Wien spectrum will have
a temperature that is different than the original temperature of the
photosphere). 

We thus conclude that broadening of the Planck spectrum naturally
occurs if the following conditions are met: (i) Dissipation processes
take place below the photosphere, at optical depth of $\tau_{\gamma e}
\sim$ few ; (ii) the energy given to the electrons is comparable to the
energy in the thermal photons component; and (iii) a strong magnetic field
exists, of the order $u_B / u_{\rm th} \approx$~tens \%. See further 
discussions in  Pe'er et al. (2005, 2006), Giannios (2006), and  Beloborodov (2010).

\subsection{Broadening of the Planck spectrum: Detailed numerical simulations}
\label{sec:asaf}

The arguments given in \S\ref{sec:Felix} above provide a guideline to
possible conditions that can lead to broadening of the thermal
spectrum. However, quantitative results can only be obtained
numerically. This is because of the non-linearity of the problem.
First, a large number of pairs can in principle be produced. Thus, a
rapid electromagnetic cascade may be presented. Second,
as most of the scatterings occur with cold electrons, the cross
section is Klein-Nishina suppressed, and hence simple analytical
approximations to the resulting spectra are absent. Finally, photons
and electrons can participate in a large number of processes, such as
synchrotron, synchrotron self-absorption and Compton scattering, which
can have similar importance.
 
In order to obtain numerical results, we use the code developed by
Pe'er \& Waxman (2005) for the study of GRB prompt emission.  This
code was further modified to the study of photospheric emission by
Pe'er et al. (2005, 2006). The numerical code solves self-consistently
the kinetic equations that describe a large number of physical
processes (synchrotron emission, synchrotron self-absorption, direct
and inverse Compton scattering, pair production and annihilation and
the development of an electromagnetic cascade) that can take place
following the injection of energetic particles close to the
photosphere. Its unique integrator enables it to follow the evolution
of the particle distribution and spectra over many orders of magnitude
in time and hence in energy scales. Thus, it is ideal for studying
processes that can take place in regions of high optical depth, in
which the characteristic time scale for interactions is much shorter
than the dynamical time. A full description of the code appears in
Pe'er \& Waxman (2005).

In the scenario considered in our calculations, the exact nature of the dissipation
process is not specified. We assume that the outflow is characterised
by steady luminosity $L_0$, and constant Lorentz factor, $\Gamma$. The
dissipation is assumed to take place at radius $r_i$, and dissipate
some fraction $\epsilon_d$ of the kinetic energy.  A fraction
$\epsilon_e$ of this energy is used to accelerate electrons while a
fraction $\epsilon_B$ is channeled into magnetic fields.


The thermal component is considered as a constant source of thermal
photons, which irradiate the interaction region during the whole
calculation, and interact with the accelerated electrons. After the
energy has been dissipated the evolution of the electron/positron
populations and the photon population is followed during one dynamical
time, $t_{dyn} = r_i / \Gamma c$, subject to synchrotron radiation,
Compton scattering and pair production and annihilation, as well as
the constant influx of thermal photons. At the end of the dynamical
time all the radiation is assumed to be released, and the emerging
spectrum is given by the photon distribution at that time. {
Since $\tau \gta 1$  at the dissipation site, one should take into account
adiabatic expansion until the photons are released at $\tau \sim 1$. 
However, since the optical depth is not more than a few, adiabatic expansion 
only marginally affects the spectrum and can be
neglected for our purposes (see further Pe'er \& Waxman 2004).}

The code allows us to quantitatively confirm the analytical results
discussed in \S \ref{sec:Felix}.  For instance, while Comptonisation naturally leads
to a harder high-energy power-law, we find that  the optical depth has to be 
relatively high for the number of scatterings to be large enough to
 affect the high-energy slope. On the other hand, even larger values of $\tau$ leads instead 
to a steepening of  $\beta$ due to thermalisation.
Moreover,  the slopes of the low-energy power-law index, $\alpha$, depends
most strongly (for a individual dissipation scenario) on the strength
of the magnetic field generated, i.e. on $\epsilon_B$, giving rise to synchrotron 
emission. A larger $\epsilon_B$ leads to a softer value of $\alpha$.
However, the effect of Comptonisation and optical depth counteract this softening. Therefore, 
the broadest photospheric spectra are obtained  for a strong dissipation occurring at
moderate optical depths, typically $\tau \sim 10$. A more detailed  accounting of the 
effects of subphotospheric heating  on the photospheric spectrum, using this code, is given 
in Nymark et al. (2011).

If shocks occur above the baryonic photosphere ($\tau << 1$) we find that the original Planck function will only be marginally affected by the shocked electrons due to the low number of scatterings that will occur. 

Note that in these calculations we consider only one dissipation
episode. In reality several dissipations are expected, making the
emerging spectrum a superposition of the spectra from several
dissipation episodes (see e.g. Giannios 2008, Pe'er et al. 2011). However, for the
purposes of this study, the spectrum resulting from one dissipation
episode is sufficient to get an indication of the effects that
dissipation can have.

\section{Application to  the spectral evolution in GRB 090902B}
\label{sec:4}

\subsection{Epoch 1}
 
In the data analysis above (\S \ref{sec:bandfunc}) we found that the averaged value of the  low-energy slope is harder than $\alpha = 0$ during epoch 1 in GRB090902B. Apart from being a constraint on the radiation process, this observation sets  constraints on any dissipation process   that can have had an influence on  the spectral shape, such as constraints on the dissipation fractions of the kinetic energy and dissipation radius.  

It is worth to note here, though,  that when observing a GRB  photosphere we actually expect to observe a superposition of spectra with different temperatures as measured in the observer frame, due to geometrical effects (Pe'er 2008). The result is a slightly broadened spectrum. Therefore the sharpest spectrum is not a Planck function but rather a multi-color blackbody (Pe'er \& Ryde 2011, see further Lundman et al. 2011).  This effect was proposed to explain the deviations from a Planck function that is observed during epoch 1 (Ryde et al. 2010, see also Larsson et al. 2010).

In any case, the simulations of subphotospheric shocks and their effect on the emitted photospheric photon energies, described in \S \ref{sec:asaf}, show that dissipation occurring at a low optical-depth only marginally affects the emitted photospheric emission. Moreover,  in the case of dissipation at  moderate optical depths, say $\tau = 10$,  a small deviation from a Planck function  
can only be achieved if  the dissipation  is relatively weak. Indeed, such dissipation can easily reproduce the observed spectral shapes during epoch 1. As an example we show in the left-hand panel of Figure \ref{fig:6}  a  spectrum found from simulating a shock dissipation in a flow  at  $\tau =  10$, and with the parameters $\epsilon_d = 0.1$, $\epsilon_e = 0.1$, $\epsilon_B = 0.1$. The peak in this spectrum is slightly broader than a Planck function (indicated in the figure with a red dashed line). This shape of the spectrum indeed resembled the MeV peak observed in GRB090902B, e.g. see the red spectrum in the lower right panel of Figure \ref{fig:1}.

Since the observed spectra do not differ much from a Planck spectrum
during this epoch, one can thus argue  that there cannot have
been significant subphotospheric {dissipation}  during this
period.  Only weak dissipation or no dissipation at all should have
occurred. The thermal peak is therefore mainly the original
non-processed thermal emission, which was formed at ({or close
  to}) the base of the flow, at very high optical depths.

 The thermal emission that is observed can either be attributed  to a baryonic photosphere or to a pair photosphere. Depending on which photosphere is assumed,  a different conclusion will be drawn. 

 \subsubsection{Baryonic photosphere}
  \label{sec:baryons}

The photospheric radius deduced from the observations to $r_{\rm ph} \sim 10^{12} $ cm can be assumed to be associated to the opacity of the baryonic electrons. This is what is assumed in Ryde et al. (2010) and Pe'er et al. (2010). Furthermore, since the photosphere can be assumed to be  non-dissipative one can use the standard theory discussed in Pe'er et al (2007) and derive that  $r_0 = 10^9 \, Y^{-3/2}  \mathrm{cm} \sim 7.6 \times 10^{8} \mathrm{cm}$  (using the time average values for GRB090902B). The value of $r_0$ implies an $ \tilde{ \alpha} \sim 3.3 \times 10^{2} \, Y^{-3/2} M_1^{-1} \sim 2.5 \times 10^{2} M_1 ^{-1} $, with the notation  $M_1 \equiv  (M/10 M_\odot)$. According to equation (\ref{eq:shock}) the large value of $\tilde{\alpha}$, and thereby $r_0$, implies that the dominant fraction of internal shocks occur well outside of the photosphere, assuming typical size of the opening angle $\theta_{\rm j}$. This is consistent with the assumption of a non-dissipative outflow.
 
 \subsubsection{Pair photosphere}
 \label{sec:pairs}
 
 The value of the  photospheric radius, $\sim 10^{12}$ cm, that  is deduced from the observations can instead be attributed to the 
 the pair photosphere.  Thus setting $r_{\rm \pm} \sim 10^{12} \, \mathrm{cm}$ then equation (\ref{eq:phpm}) yields\footnote{Using $\Gamma \sim 1000$, which is inferred from opacity arguments (Abdo et al. 2009), $r_{\rm ph} \sim 2 \times 10^{12}$ cm and $\epsilon_{\pm}  \, \,  \tilde{ \alpha} ^{-1}  \sim 5 \times 10^{-3}$} that
$\epsilon_{\pm}  \, \,  \tilde{ \alpha} ^{-1} =  1.25 \times 10^{-3} L_{53}^{-1} \Gamma_{2.8}^3$.

Furthermore, {equations (\ref{eq:ph}) and (\ref{eq:phpm})  imply that the ratio of the photospheres is  ${r_{\pm}}/{r_{\rm ph}}  
 \sim 1.8 \times 10^{3} \, \epsilon_\pm \,  \tilde{ \alpha} ^{-1}$ 
in the coasting phase.}
Thus  ${r_{\pm}}/{r_{\rm ph}} = 2.1 \,L_{53}^{-1} \Gamma_{2.8}^3 $ and thereby the baryon-photospheric radius is ${r_{\rm ph}}  \sim 5.0 \times 10^{11} \, L_{53}  \Gamma_{2.8}^{-3}$ cm.
Furthermore, $r_0 = \tilde{\alpha} r_{\rm g} = 2.4~\times~10^{9} \epsilon_\pm 
M_1 L_{53} \Gamma_{2.8}^{-3} \mathrm{cm}$ and the saturation radius $r_{\rm s} = \Gamma r_0 = 1.5~\times~10^{12} \epsilon_\pm  M_1 L_{53} \Gamma_{2.8}^{-2} \mathrm{cm}$.

{Pairs are either created at the base of the flow (Goodman 1986)  or in the energy dissipation at the shock.}
The requirement that $r_{\rm sh} \lta r_\pm$ (pairs are created in the shock), with $r_{\rm sh}$ given by equation (\ref{eq:shock}), then corresponds to 
$ \tilde{\alpha} \lta 1.6$. 
From the estimate of $\epsilon_{\pm}  \, \, \tilde{ \alpha} ^{-1}$ above this is equivalent of $ \epsilon_\pm <  1.58 \times 10^{-3}$. This value implies that in this scenario $r_0 \sim 3 \times 10^{6} \rm{cm}$ and $r_{\rm s} \sim 2 \times 10^{9} \rm{cm}$ for GRB090902B.
{ This implies that the characteristic initial temperature of the fireball is of the order  $T_0 \sim 4$ MeV (e.g. Rees \& M\'esz\'aros 2005). Therefore, the expected observed temperature  $T = T_0 (r_\pm / r_{\rm s})^{-2/3}$ is approximately $100$ keV due to adiabatic expansion. We note that this is a factor of a few below the measured peak energy values.  However, the assumption of adiabatic expansion is not necessarily valid since energy dissipation occurs below the photosphere to create the pairs.  As discussed in \S 3.1 the heating is most probably continuous due to oblique shocks.}

Assuming that the photospheric emission that we are  observing is due to the pair photosphere thus results in  a value of $r_0$ which is similar to the generally assumed value for the jet bounding radius. It naturally alleviates the concern that the value, found for the baryonic photosphere ($r_0 \sim 10^8$ cm), is too large. It can thus be argued that it is indeed the pair photosphere we are observing. We note, however, that  Zhang \& Woosley (2004) found typical values close to $r_0 \gta 10^{8}$ cm,  since the jet is not well collimated at the center thus preventing the acceleration. This value is similar to what we deduce for the baryonic photosphere.




\begin{figure*}
\begin{center}
\resizebox{84mm}{!}{\includegraphics{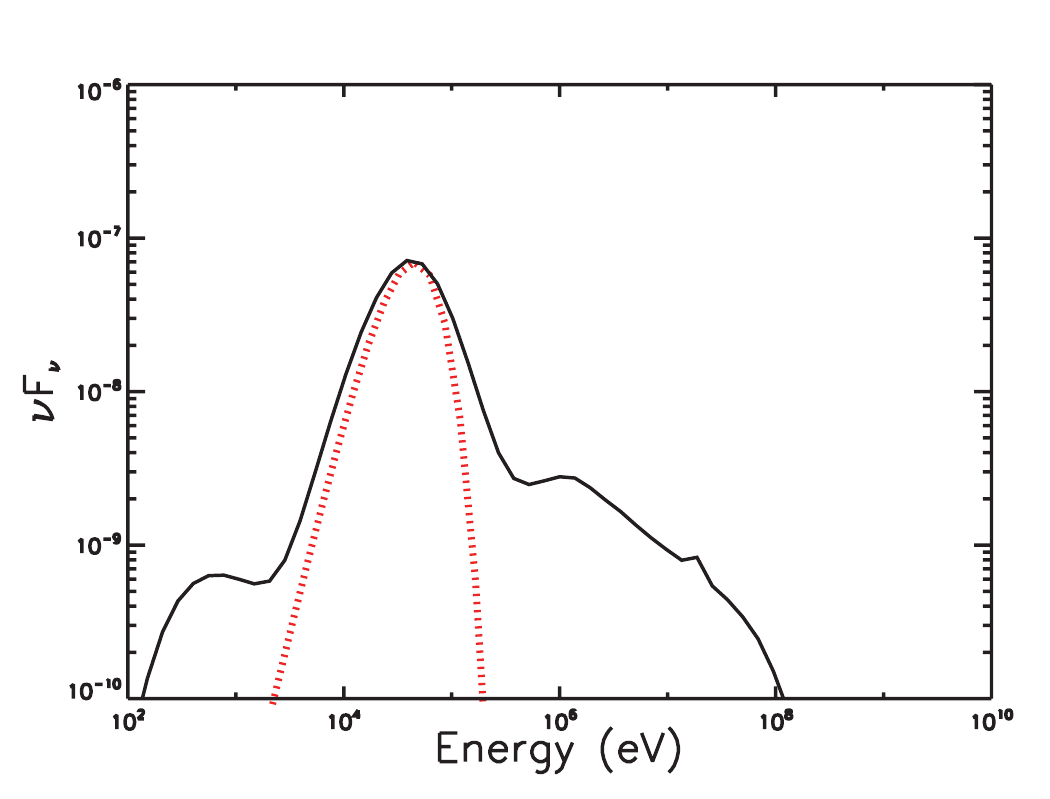}}
\resizebox{86mm}{!}{\includegraphics{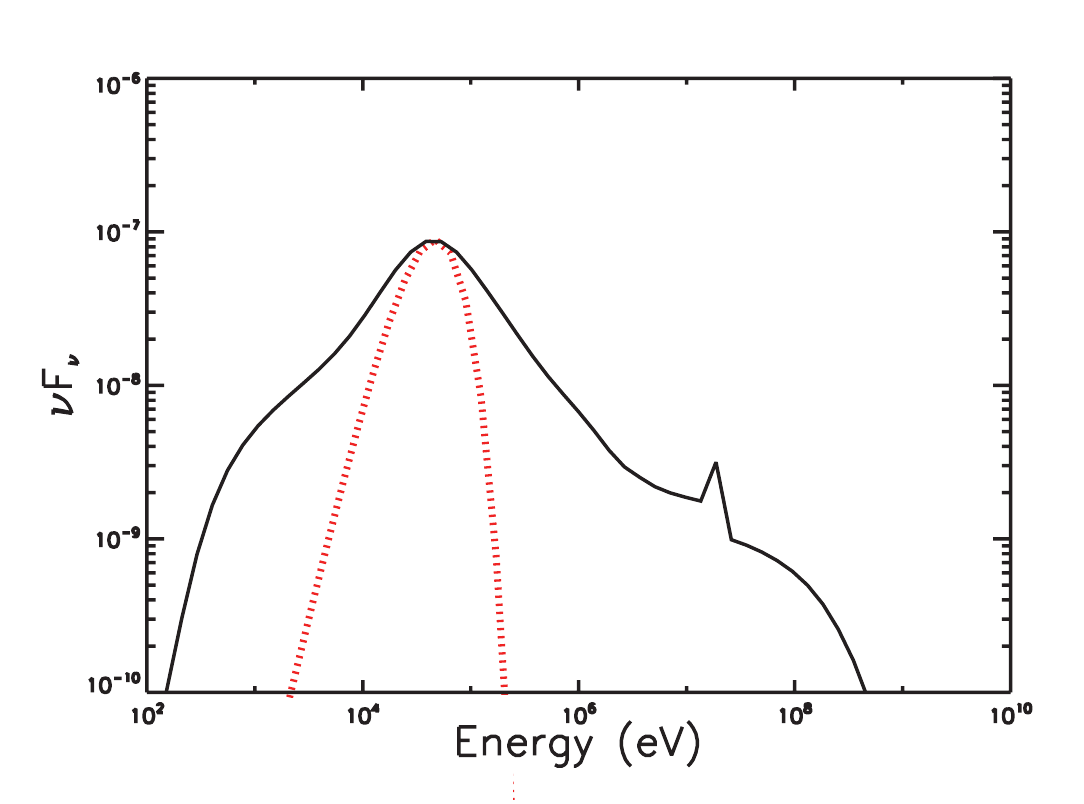}}
\caption{\small{Generic model spectra from subphotospheric shock heating (solid line), illustrating the broadening effect. {\it Left panel.} A low level of dissipation only slightly distorts the MeV peak from begin a Planck function. This spectrum can explain the observed shape at 6.5 s (see Fig. \ref{fig:1}). {\it Right Panel.} A higher level of dissipation broadens the photospheric  peak, leading to a Band-like spectrum. This spectrum can explain the spectrum at 22 s (see Fig. \ref{fig:1}). The dashed, red line shows the shape of a Planck spectrum at the temperature that corresponds to the peak of the spectrum.}}
\label{fig:6}
\end{center}
\end{figure*}

\subsection{Epoch 2}

During epoch 2 strong dissipation has to occur at a moderate optical
depth, $\tau \sim$ few,
in order to broaden the MeV bump. 
{From simulating various dissipation scenarios (described in \S 3.3),  we conclude, in particular, that the low-energy slope is mainly determined by the contribution of synchrotron emission ($\epsilon_B$): For instance, a power-law distribution of electrons, produced in the dissipation process, is expected to have a peak energy at $\gamma_m \sim \epsilon_e (m_p/m_e) = 1860 \epsilon _e$. For the typical parameters for GRRB090902B this translates into a  peak of the synchrotron spectrum at $E^{synch}_m = 186 \, \, L_{53}^{1/2} \epsilon_{b,-1}^{1/2} \epsilon_{e,-1} R^{-1}_{12}$ keV, which is less than the averages thermal peak lying at $E_p^{\rm thermal} = 2.82  <kT> \sim 840$ keV.}  This is still  the case if  $\epsilon_{b}$ and $ \epsilon_{e}$ are somewhat larger, approaching the equipartition values.

The right-hand panel in Figure \ref{fig:6} shows an example spectrum from a simulation of a shock dissipation at the same optical depth as before ($\tau =  10$), but with an increased energy dissipation. In this example, the dissipation fractions are given by $\epsilon_d = 0.2$, $\epsilon_e = 0.3$, $\epsilon_B = 0.3$. 
This spectrum is similar in shape to the blue spectrum in Figure \ref{fig:1}, illustrating that subphotospheric heating indeed  can explain the change in the spectral shape that is observed in epoch 2.

\subsection{Transition between the two epochs}
\label{sec:transition}

Variations at the base of the flow are expected to lead to rapid variation of $\Gamma$ and $r_{\rm ph}$ of the photosphere, down to a time-scale of $r_0 \theta_{\rm j}/c$ (Rees \&  M\'esz\'aros 2005). This causes varying properties of the dissipation, such as its strength and where it occurs relative to the photosphere.

Due to strong dependence on the Lorentz factor of the ratio ${r_{\rm sh}}/{r_{\rm ph}} \propto \Gamma ^{5}$ a change in $\Gamma$ (apart from variation in the other parameters) can easily alter the dissipation pattern and thereby the appearance of the photospheric spectrum.  A decrease in $\Gamma$ is, for instance, able to cause $r_{\rm ph}$ to become larger than $r_{\rm sh}$, that is,  shocks mainly appear  below the  baryonic photosphere - subphotospheric shocks.   Such a change is hence a plausible explanation the change in the spectrum at $\sim 12.5$ s in GRB090902B.  Indeed,  in both the baryonic (\S \ref{sec:baryons})  and pair scenarios (\S \ref{sec:pairs}) the bulk of the dissipation site then moves from being above $r_{\rm ph}$ to being below it, mainly due to a drop in $\Gamma$.  
 
{In the photospheric model, the peak of the spectrum { (which} is measured by $E_{\rm p}$) should be closely related to the temperature of the photosphere. This  is, for instance,  illustrated by Figure \ref{fig:6}.   Furthermore, in deriving the Lorentz factor, $\Gamma$, from the observables, we note that it  most strongly depends on the temperature ($\Gamma \propto T^{1/2}$, Pe'er et al. 2007): The evolution of  $\Gamma$ and $kT$ are expected to track each other.
We therefore argue that variations in $E_{\rm p}$ are closely related to corresponding variations in $\Gamma$ (apart from the Doppler boost). In Figure \ref{fig:4} we show that there is a drop in averaged $E_{\rm p}$ between epoch 1 and epoch 2. This thus indicates that a drop occurred in the averaged temperature at this time and thereby also a drop in $\Gamma$.  

The change in $\Gamma$ leads to a change  in dissipation pattern, according to the discussion above, and thereby the spectral shape.
Indeed, Figure \ref{fig:corr} shows a correlation between $E_{\rm p}$ and $\alpha$, reinforcing this interpretation; a decrease in $E_{\rm p}$ corresponds to a decrease in $\Gamma$, which leads to more subphotospheric heating thereby broadening the spectrum (a decrease of $\alpha$).  This is most clearly illustrated by the local increase in $E_{\rm p}$ observed around 15 s (see Fig. \ref{fig:4}), which corresponds to a local increase in $\alpha$, indicating less subphotospheric heating.  }

 
\begin{figure}
\begin{center}
\rotatebox{0}{\resizebox{!}{60mm}{\includegraphics{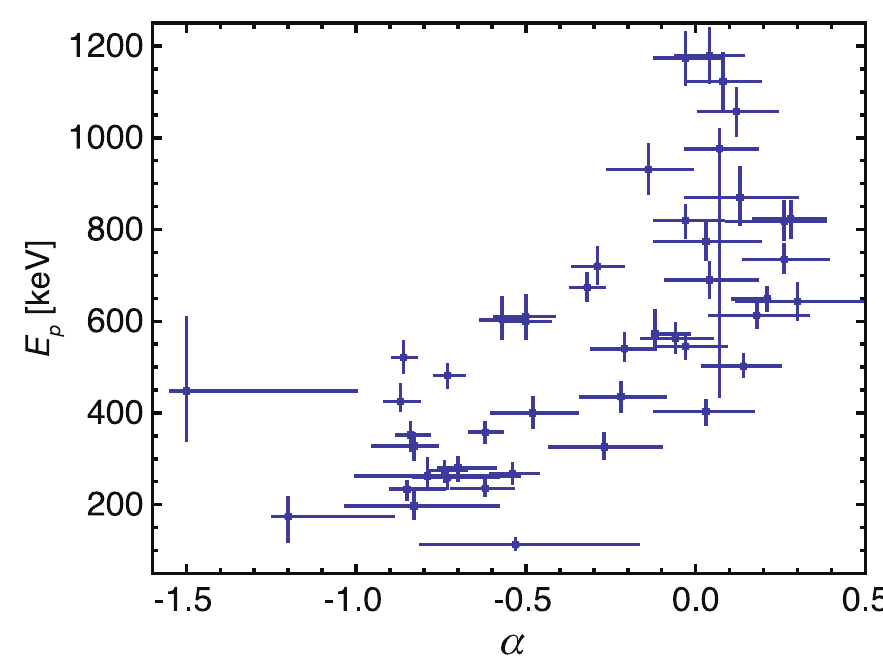}}}
\caption{\small{Correlation between 
$E_{\rm p}$ and $\alpha$. Interpreting $E_{\rm p}$ as related to the temperature of the photosphere, this can be explained by the expected relation between temperature and spectral width in the presence of subphotospheric heating.}}
\label{fig:corr}
\end{center}
\end{figure}

\section{Discussion}

The remarkable property of GRB 090902B is that it is dominated, at early times, by a very peaked spectral component in addition to a power law component. This peaked component, which has a spectral shape very close to a Planck function, can only be  explained by emission from the photosphere.  As we have seen in the analysis  above, at late times the spectrum of this peaked component is broadened into a spectrum that can be described 
by a Band function with an $\alpha$ slope of approximately -0.6 and  a $\beta$ slope of approximately -2.5. We note that this is close to the spectral shape usually attributed to GRBs (Kaneko et al. 2006). Since there is a continuous change in shape, even though it occurs over a short period compared to the burst duration, it has to be concluded that the photosphere emission continues during the second half of the burst. The consequence of this conclusion is that the photosphere emission spectrum can have a large variety of shapes, which reflect the dissipation processes in the flow. 
We have argued that this burst {provides} observational evidence for subphotospheric heating.

This mechanism  provides a natural explanation to the observed variety 
of spectral shapes in GRBs. In addition, the more typical spectral evolution,
in which the spectra gradually become softer, can be explained by a gradual change in the dissipation pattern in the flow. 

The shape of the photospheric peak, given, for instance, by $\alpha$ and $\beta$ of a Band function fit, can now be translated into physical properties of the dissipation, quantified by the parameters such as $\epsilon_e$, $\epsilon_B$, and $\tau$ and  dissipation rates. Depending on the details of the model  we can therefore diagnose the outflow and its temporal evolution in individual bursts.


\subsection{Photospheric emission}
\label{sec:51}

Among the main contenders for explaining the hard $\alpha$ values observed are (i) small-pitch-angle synchrotron emission (Epstein 1973; Epstein \& Petrosian 1973) or similarly jitter radiation (Medvedev 2000, 2006), and (ii) inverse Compton emission seeded by self-absorbed synchrotron emission (Painatescu \& M\'esz\'aros 2000) or by soft photons with a narrow energy distribution, i.e., a quasi-monoenergetic distribution (Stern \& Poutanen 2004).  However, these non-thermal emission models typically lead to very broad spectral peaks, and  cannot produce spectra that are as narrow as observed (below an order of magnitude) and that are as  hard ($\alpha \gta 0$, see \S \ref{sec:alpha0}).  
For instance, inverse Compton emission leads to broad spectral peaks of  typically 2 orders of magnitude or above  (see, e.g. Stern \& Poutanen 2004, Baring \& Braby 2004).

Moreover, studies of the acceleration processes in relativistic collisionless shocks indicate that  a strong thermal component in the electron spectrum is formed (Sironi \& Spitkovski 2010). Even though the full emitting region can still not be fully simulated, these results question the classical assumption of a strong power-law component of the shocked electron population.
Sironi \& Spitkovski (2010) show through particle-in-cell simulations of  shocks in unmagnetised pair plasmas  that the width of the emitted synchrotron spectrum, in general, lie in the range of 2 -- 4 orders of magnitude.

We further note that Sironi \& Spitkovski (2010) argue that  if the electrons are  accelerated in relativistic unmagnetised shocks then the emission is in the classical synchrotron regime rather than in the jitter radiation regime. In addition, small-pitch-angle synchrotron emission predicts a negative correlation between $E_{\rm p}$ and $\alpha$, in contrast to the observed (mainly) positive correlations (Lloyd-Ronning \& Petrosian 2002). These two facts pose further challenges for such an explanation for the hard spectra observed.

The narrow and hard spectra we observe  in GRB090902B are thus inconsistent with {the} non-thermal emission models. 

The photospheric model can easily overcome many of the challenges of the standard, internal-shock synchrotron model (see, e.g., discussion in Ryde \& Pe'er 2009).  Most importantly, the (reprocessed) Planck function naturally provides very hard spectral slopes ($\alpha \leq 1 $). Moreover, a power-law distribution of electrons is not required to be produced by the acceleration processes; the power-law slope in the BB+pl model has a preferred value of $\alpha \sim -1.5$ (Battelino et al. 2002, Ryde  \& Pe'er 2009). This is naturally expected due to the cooling of the electrons, which produces a power-law distribution below the characteristic synchrotron frequency (the case is similar for SSC). 

An inevitable and characteristic signature of emission from the photosphere is a cut-off at energies $0.5\, \Gamma \,   \rm{MeV}  \gta 100$ MeV due to pair production. The observation of an extension of a Band function to energies above several GeV in bursts (e.g. GRB080916c, Abdo et al. 2009) may thus pose a challenge for (reprocessed) photospheric emission to explain the spectra.
In order to overcome this difficulty, the existence of very  high-energy photons in the observed spectra therefore requires an additional emission site, which is capable of producing optically-thin emission. We note that such a scenario (spectra indicating two emission components) is clearly observed in several bursts observed by Fermi-LAT. Moreover, in one of the analysed timebins in GRB080916c there is indeed an indication of an extra power-law component, even though the overall conclusion of the analysis is that a single component sufficiently fits that data (Abdo et al. 2009). This suggests the possibility that an extra component could indeed be present in the  data. However, the combined spectrum can still be satisfactorily   fitted by an extension of a single Band function, found at lower energies. Ghisellini et al. (2010) indeed find that the power-law slope fitted to the spectrum at  energies $\lta 40 $ MeV (i.e., the Band $\beta$) and the power-law slope fitted to the spectrum at energies at $\lta 100 $ MeV, are significantly different. This can be an additional indication that additional emission components are required at the highest energies, thereby alleviating this problem for the photospheric model.

\subsection{Correlation between $E_{\rm p}$ and $\alpha$}

In \S \ref{sec:transition} we discuss the observed correlation between $E_{\rm p}$ and $\alpha$ in terms of subphotospheric heating.
A positive correlation between time-resolved $E_{\rm p}$ and $\alpha$ values for individual bursts was early identified by, for instance, Ford et al. (1994), Crider et al. (1997), and Lloyd-Ronning \& Petrosian (2002). More recently, Kaneko et al. (2006) found that this is the strongest correlation among GRB parameters. In a sample of 196 bursts they found a strong correlation for 26 \%; in most cases the correlation is stronger than the one in Figure \ref{fig:corr}.
The actual fraction of bursts exhibiting a strong correlation can be even higher, since measurement uncertainties in many cases may have masked the correlation.

The fact that not all bursts exhibit a strong correlation between $E_{\rm p}$ and $\alpha$  might also be an indication of  that the observed spectral peaks are not directly related to $kT$.  In some bursts a weak contribution of a Planck function on top of a dominant  Band spectrum can be identified (Ryde \& Pe'er 2009, Guiriec et al. 2011). The power peaks are, in these cases, given by the non-thermal Band function and should thus not be identified by $kT$.
As discussed in \S \ref{sec:Felix} this could be the case  if the energy density of the electrons  is much larger than the energy density of the 
thermal photons. Then the thermal component is expected to be relatively weak.

\subsection{Change in spectral shape from "thermal" to "non-thermal"}

Other scenarios than the one described in \S \ref{sec:transition} can be envisioned to describe a change from
quasi-Planck spectrum to Band-type spectrum. 
For instance, Beloborodov (2010)
calculated the full radiative transfer of the relativistic jet and
showed that a perfect Planck function is obtained if and only if  the
flow is dominated by radiation, that is, the radiation energy density
is much larger than the rest mass energy density of the plasma  in the
comoving frame of the flow. This is independent of subphotospheric
dissipation since the photon energy totally dominates. However, there will inevitably {be} broadening
due to geometrical effects (Pe'er \& Ryde 2011), as discussed in  \S\ref{sec:Felix}.
 The spectral change observed between epoch 1 and 2 can then be due to a change from a radiation dominated phase into a baryon dominated phase, in which dissipation  causes the strong deviation from the Planck function. An interesting consequence of such a scenario would be vanishing of the strong polarisation expected during the photon dominated phase.

Yet another possibility is if the variability time-scale and amplitude suddenly change.  During epoch 2 the variations of the temperature can become large and be on a time scale smaller than the integration time scale. We would then measure spectra that are broader than a Planck spectrum due to significant variations of the spectral peak during the integration time. In such a case the asymptotic slopes of the spectrum should still be those of a blackbody.

Finally, one may still envision that the emission in the MeV peak during epoch 1 is due to the photosphere while during epoch 2 the emission is from a different, non-thermal, radiation process, for instance, from optically-thin synchrotron emission. However, as discussed above the observed spectra during epoch 2 are still significantly inconsistent with what is expected from the fast cooling electrons ($\alpha =-1.5$ and a much broader spectral width). Moreover, to get the synchrotron peak-energy to lie in a similar energy range as the thermal peak during epoch 1 requires an unreasonable coincidence. This is also not supported by the behavior of the NT part of the spectrum.

\subsection{Comparison with Pre-Fermi Spectral Analysis}

Prior to the launch of Fermi we mainly had to rely on spectra in a narrow energy range, for instance that of {\it CGRO}-BATSE  ($\sim 25 - 1900$ keV). The limited spectral widths made it difficult to unambiguously deduce the spectral {behaviour} of GRBs. Nevertheless, the importance of the photospheric emission was already alluded to (see, e.g., the review by Ryde 2008).   Using the broader energy range provided by Fermi this unambiguity can now be {revealed}.  We can summarise the pre-Fermi results with the following three behaviours.

(i) Many bursts provided us with a clear indication of the photosphere. These include the spectra that are well described by a single Planck function (Ryde 2004) and the spectra for which a fit of the BB+pl model gives a statistically significant  improvement over a fit with a Band function model (Ryde \& Pe'er 2009).  Such cases should be similar to GRB090902B observed by Fermi; the $\nu F_{\nu}$ peak is due to the Planck function and there are two distinct components.

(ii) Other bursts indicated  that the power-peak lay beyond the observed energy range, even though a subdominant, thermal peak was identifiable: The thermal component forms a shoulder on the low-energy side of the power-peak. The power peak of the spectrum is, in this interpretation, not directly due to the photospheric component, but due to a non-thermal emission. Example of such spectra are given by PHEBUS Granat observations of GRB 900520a (Barat et al. 1998), {\it CGRO} BATSE/EGRET observations of GRB981021;  BATSE trigger 7071 (Gonzalez et al. 2009; Ryde \& Pe'er 2009), and Fermi observations of GRB100724B (Guiriec et al. 2011), as well as Fermi cases like GRB 080916c (Abdo et al. 2009).

(iii) As argued in the study above, the thermal peak can be broadened due to subphotospheric heating, which creates a spectrum which is Band-like and significantly different from a Planck function. This mechanism is thus  able to explain many Band-like spectra as well as the typical hard-to-soft spectral evolution. Apart from these facts, it also opens up the possibility that some spectra that were fitted in pre-Fermi analyses by a BB+pl model, in a narrow energy range, may have been misinterpreted. This could be the case particularly for bursts whose BB+pl spectral fits are not an improvement over a Band-only model fit. In these cases, the Planck function component of the fits still correctly captures the photospheric peak. However, the power-law component instead captures the broadening of the peak (which makes it non-Planckian), instead of representing a real secondary component. 
If this is the case, such a BB+pl model would, of course,  not be able to describe the broad band spectrum of these bursts. Such spectra can explain the results in Ghirlanda et al. (2007) and Bellm (2009), who find that in some bursts a simple extension of the BB+pl model does not seem to fit the data at hand outside the BATSE energy range. We note that this is apart from the fact that  the non-thermal component is expected to be a more complicated function than a single power-law over a broader energy range.

\section{Conclusion}

The unambiguous signature of emission from the photosphere is a Planck function. However, subphotospheric dissipation can easily distort the photospheric emission into a broader spectrum, resembling a Band function.

The burst of 090902B made it possible for us to draw two important conclusions. First, the study of the behaviour of the MeV peak
allows us to observationally claim that  the origin of a Band function can be the same as the origin of a Planck  spectrum; a Band spectrum can  be interpreted as photospheric emission. Second, in order to explain the broader spectral shape of the photospheric emission, subphotospheric dissipation is needed. The existence of such dissipation is thus verified.

We show that a varying Lorentz factor of the outflow leads to a varying strength of subphotospheric dissipation. This, in its turn, leads naturally  to a correlation between the broadening of the spectrum and its peak energy. A correlation between $\alpha$ and $E_{\rm p}$, which is a consequence of this, is indeed commonly observed in GRBs. 

Photospheric emission could thus be a ubiquitous signature of the prompt emission spectrum and play an important role in creating the diverse spectral shapes and spectral evolutions that are observed.
The photospheric component can be dominant (e.g. GRB090902B) or subdominant (e.g. GRB100724B). It can be a Planck function throughout the burst (Ryde 2004) or it can broaden with time (as argued in this paper).  An important consequence of this is that broad-band fitting of GRB spectra with only a Band function might be misleading.

\vskip 4mm
{\bf ACKNOWLEDGEMENTS}
\vskip 3mm
We thank Peter M\'esz\'aros for useful comments on the manuscript.
FR wishes to express his gratitude to the Swedish National Space Board for financial support.
EM acknowledges with thanks financial support from Carl Tryggers Stiftelse f\"or Vetenskaplig Forskning. 

The \textit{Fermi} LAT Collaboration acknowledges generous ongoing
support from a number of agencies and institutes that have supported
both the development and the operation of the LAT as well as
scientific data analysis.  These include the National Aeronautics and
Space Administration and the Department of Energy in the United
States, the Commissariat \`a l'Energie Atomique and the Centre
National de la Recherche Scientifique / Institut National de Physique
Nucl\'eaire et de Physique des Particules in France, the Agenzia
Spaziale Italiana and the Istituto Nazionale di Fisica Nucleare in
Italy, the Ministry of Education, Culture, Sports, Science and
Technology (MEXT), High Energy Accelerator Research Organization (KEK)
and Japan Aerospace Exploration Agency (JAXA) in Japan, and the
K.~A.~Wallenberg Foundation, the Swedish Research Council and the
Swedish National Space Board in Sweden.

Additional support for science analysis during the operations phase is
gratefully acknowledged from the Istituto Nazionale di Astrofisica in
Italy and the Centre National d'\'Etudes Spatiales in France.

The Fermi GBM collaboration acknowledges support for GBM development,
operations and data analysis from NASA in the US and BMWi/DLR in
Germany.

\end{document}